\documentstyle[aps,epsfig]{revtex}
\begin{document}
\twocolumn[\hsize\textwidth\columnwidth\hsize\csname @twocolumnfalse\endcsname

\draft
\title{
One Dimensional Kondo Lattice Model Studied by the
Density Matrix Renormalization Group Method
}

\author{Naokazu Shibata}
\address{
Institute of Applied Physics, University of Tsukuba,
Tsukuba 305, Japan
}
\author{Kazuo Ueda}
\address{
Institute for Solid State Physics, University of Tokyo,
7-22-1 Roppongi, Minato-ku, Tokyo 106-8666, Japan
}
\date{\today}

\maketitle
\begin{abstract}
Recent developments of the theoretical investigations on 
the one-dimensional Kondo lattice model  
by using the density matrix renormalization group (DMRG) 
method are discussed in this review.  
Short summaries are given for the zero-temperature DMRG, 
the finite-temperature 
DMRG, and also its application to dynamic quantities.

Away from half-filling, the paramagnetic metallic state is shown to be a 
Tomonaga-Luttinger liquid with the large Fermi surface. For the large Fermi
surface its size is determined by the sum of the densities of
the conduction electrons and the localized spins.  
The correlation exponent $K_\rho$ of this metallic phase is 
smaller than $1/2$.
At half-filling the ground state is insulating. Excitation gaps are
different depending on channels, the spin gap, the charge gap and the
quasiparticle gap. 
Temperature dependence of the spin and charge 
susceptibilities and specific heat are discussed.
Particularly interesting is the temperature dependence of various excitation 
spectra, which show unusual properties of the Kondo 
insulators. 
\end{abstract}

\vskip2pc]

\narrowtext
%
%
%
%
%
%

\section{Introduction}

In a degenerate Fermi gas, low temperature specific heat is
linear in $T$ and the proportionality constant is given by the
density of states at the Fermi energy.  For the free electrons 
the density of states including the two spin directions is given by 
\begin{equation}
   D(\epsilon_{\rm F}) = \frac{mk_{\rm F}}{\pi^2\hbar^2}
\end{equation}
where $m$ is the free electron mass and $k_{\rm F}$ is the Fermi 
wave number. 

To include the effect of electron-electron interaction, Landau
developed the theory of Fermi liquids.  Since the volume of 
the Fermi sphere is determined by the density of electrons, it 
is a reasonable assumption that the Fermi wave number does not change by 
the interaction, which was in fact proven later by Luttinger
\cite{Luttinger}.
According to Landau, the effect of the interaction can be taken 
into account by replacing the bare electron mass $m$ by an effective
mass $m^*$. Thus importance of the interaction effects in each 
material may be judged from the electronic specific heat at low 
temperatures. 

In ordinary metals the coefficient of the $T$-linear term, $\gamma$, 
is of the order of mJ/mol K$^2$.  However, in some rare earth and 
actinide compounds there is a group of compounds whose $\gamma$ are 
in the range from 0.1 J/mol K$^2$ to more than 1 J/mol K$^2$.  This 
class of materials are called heavy Fermion systems or heavy electron 
systems. A key feature of the heavy Fermion systems is that it
contains two different types of electrons: relatively localized 
$f$ electrons and extended conduction electrons. Interplay between
the two degrees of freedom is an essence of the heavy Fermion physics.

Since the Coulomb interaction between the $f$ electrons is strong, 
a partially filled $f$ shell in an isolated ion possesses a well defined 
magnetic moment corresponding  
to the total angular momentum of the $f$ shell.  A weak 
hybridization between the $f$ electrons and the conduction electrons
is the source of interesting manybody problems.  

When we consider a
single $f$ shell in the sea of conduction electrons, the magnetic 
moment of the $f$ electrons is unstable, leading to the Kondo singlet
which is a bound state of the $f$ moment with the spin polarization 
of the conduction electrons\cite{Yosida}.  
When we consider two $f$ shells, spin polarization of the
conduction electrons induced by an $f$ moment tends to stabilize
the magnetic moment of the other $f$ shell.  This is the origin
of the Ruderman-Kittel-Kasuya-Yosida (RKKY) interaction\cite{RKKY}.  
Thus the Kondo effect and the RKKY interaction in many cases compete 
with each other.

If the Kondo effect dominates over the RKKY interaction by some reason, 
a paramagnetic heavy Fermion state will be stabilized.  In this regime, 
the Kondo temperature or the effective Fermi temperature in a lattice 
problem sets a small energy scale at low temperatures. Existence of
the small energy scale naturally leads to the large specific heat since 
the entropy associated with the magnetic degrees of freedom of $f$ orbitals 
should be released in the small temperature range.  

The most simple theoretical model for the heavy Fermion physics is 
the Kondo lattice model. The Kondo lattice model is given by
\begin{equation}
  {\cal H}= \sum_{\langle ij \rangle} \sum_{s} t_{ij}c^{\dagger}_{is}c_{js}
  +J\sum_{i}\sum_{s,s'}\vec{S}_{i}\cdot\frac{1}{2}\vec{\sigma}_{ss'}
  c^{\dagger}_{is}c_{is'}
\label{KLM}
\end{equation}
where $\vec{\sigma}=(\sigma_x,\sigma_y,\sigma_z)$ are the Pauli matrices
and $\vec{S}_{i}=\sum_{s,s'}\frac{1}{2}\vec{\sigma}_{ss'}
f^{\dagger}_{is}f_{is'}$
is the $f$-electron spin at site $i$.  In this review we will consider 
the model with only nearest neighbor hoppings, $t_{ij}=-t$ for the nearest 
neighbor pairs.

\begin{figure}
  \epsfxsize=85mm \epsffile{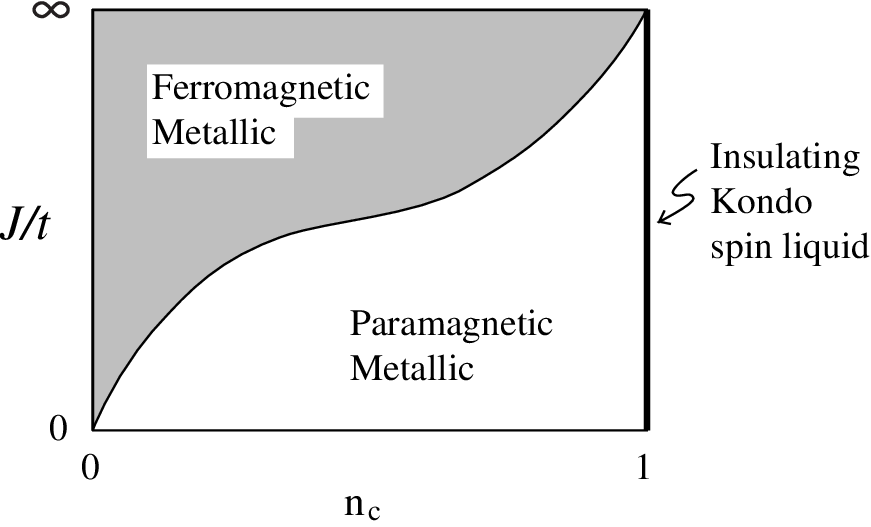}
\caption{
The ground-state phase diagram of the one dimensional Kondo lattice
model with the nearest neighbor hoppings.
}
\label{phase-dgm}
\end{figure}

Much effort has been invested for the study of the model and a significant 
progress has been achieved in one dimension in the last ten years \cite{RMP}. 
When we fix a lattice structure, the Kondo lattice model has only 
two parameters:
one is the density of conduction electrons $n_c$ and the other is 
the strength of the exchange coupling normalized by the hopping energy
$J/t$.  In one dimension, the ground-state phase diagram in the 
parameter space is completed and shown in Fig. \ref{phase-dgm}.

There are three different phases in the phase diagram. 
In the region from the low density limit to the 
strong coupling limit, a ferromagnetic metallic phase is stabilized.
The spin quantum number in this phase is given by $S_{\rm tot}=(L -
N_c)/2$, where $L$ is the number of lattice sites and $N_c$ the
number of conduction electrons. Note that the magnetic moment
vanishes as the half-filling is approached. The line of half-filling 
is special in the sense that the ground state is always a nonmagnetic
insulator.
Since the lowest excitation in this phase is a spin-triplet excitation
with a finite excitation gap,
this phase is called an incompressible spin liquid phase.  
In the remaining part of the phase diagram (Fig.\ref{phase-dgm}) 
which extends from the weak coupling limit
towards the line of half-filling, the ground state is metallic and
paramagnetic.  In this review we will discuss properties of the 
spin liquid phase and the paramagnetic metallic phase. 

Existence of the small energy scale in the Kondo lattice model means 
that correlation lengths are generally long. For example, to study 
the spin gap and the charge gap in the spin liquid phase the exact 
diagonalization was used at first \cite{THUS}.  
However the largest system size which 
can be diagonalized by the Lanczos algorithm is just ten sites. 
This is the reason why a nontrivial form of a finite size scaling 
was necessary to obtain the functional form of the spin gap.  
 
Recently Steven White developed the density matrix renormalization group 
(DMRG) method to study the ground-state properties of one dimensional 
many body systems \cite{DMRG}.  Advantage of this method is that
an order of magnitude bigger systems can be studied 
compared with the numerical exact diagonalization 
by the Lanczos algorithm. Compared with quantum Monte Carlo
simulations the DMRG is free from statistical errors.  The numerical
errors in the DMRG come from truncation errors but they can be 
estimated from the largest eigenvalue of the density matrix which 
is truncated out.  The truncation error may be improved by increasing 
the number of basis states for the density matrix. 
The DMRG method is an ideal tool to study the 
one-dimensional Kondo lattice model and in this review  we will
discuss recent developments on this subject.

Further development of the DMRG method was achieved last year 
when one of the authors \cite{shibata}
and Wang and Xiang\cite{wang} independently succeeded to  
obtain thermodynamic properties of the one-dimensional quantum XXZ 
model by applying the DMRG to the transfer matrix (finite-$T$ DMRG).
Application of the finite-$T$ DMRG to a system with Fermion degrees 
of freedom started from the Kondo lattice model \cite{SATSU}.

The present article is organized as follows. In the next section,
after a brief summary of the DMRG method for the ground state, 
we will describe the method to calculate thermodynamic properties 
by the finite-$T$ DMRG and then extend discussions to the 
dynamic quantities at finite temperatures.  In Section {\bf III}
nature of the paramagnetic metallic phase away from half-filling 
is shown to be a Tomonaga-Luttinger liquid with a large Fermi 
surface.  The large Fermi surface means that the volume inside 
the Fermi surface is determined not only by the density of 
conduction electrons but also includes the localized spins. 
Section {\bf IV} is devoted to the discussion of the spin liquid
phase at half-filling.  After the discussion of the spin gap and 
the charge gap at zero temperature, we will discuss how the excitation 
gaps develop as the temperature is lowered.  We will conclude the
present review by summary and discussions in Section {\bf V}.
  
\section{The Density Matrix Renormalization Group Method}

The density matrix renormalization group (DMRG) method is 
relatively new \cite{DMRG} among the various numerical algorithms
to treat many-body problems. However it is now widely used as  
one of the most standard numerical methods for low dimensional 
many-body systems. 
In this section we first briefly outline the algorithm of the 
zero-temperature DMRG that was developed to study
ground-state and low energy excitations of one-dimensional systems. 
The application of this method to the quantum transfer matrix enables
us to obtain thermodynamic quantities \cite{shibata,wang}
and the dynamical correlation
functions at finite temperatures.  In the second part of this section
we will summarize the algorithm of the finite-$T$ DMRG. 

\subsection{Zero temperature algorithm}

\begin{figure}
  \epsfxsize=85mm \epsffile{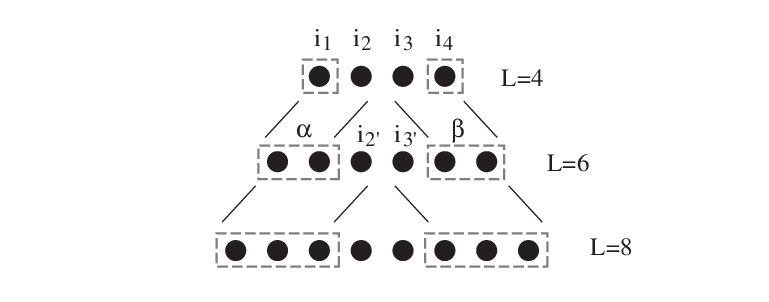}
\caption{
Schematic diagram of the infinite system algorithm of the DMRG.
}
\label{DMRG1}
\end{figure}

The zero-temperature DMRG method is designed to obtain 
the ground-state wave function and the low energy excitations 
with small systematic errors.
The ground-state wave function and the low energy excitations 
of long systems are 
obtained by expanding the system size iteratively as shown in Fig.
\ref{DMRG1}.
The expansion of the system is done by putting additional sites
in its central region to minimize undesirable boundary effects on 
the added sites.
The algorithm is described in the following.

Let us start from a system of four identical sites, for example,
a four-sites spin chain under the open boundary conditions.
An  operator on the $n$th site, e.g. $S_n$, is represented in 
terms of the complete basis states $|i_n \rangle$ as
\begin{equation}
\langle i_n |S_n| i'_n\rangle = (S_n)_{i_n,i'_n} .
\end{equation}
Then we construct a representation of the Hamiltonian 
$H_{i_1 i_2 i_3 i_4,i'_1 i'_2 i'_3 i'_4}$ for the total system.
The ground-state eigenvector 
\begin{equation}
|\Psi_{i_1 i_2 i_3 i_4}\rangle = \Psi_{i_1 i_2 i_3 i_4} |i_1 \rangle 
|i_2 \rangle |i_3 \rangle |i_4 \rangle
\label{DM_s}
\end{equation}
is obtained by diagonalizing the Hamiltonian matrix 
by some method like the Lanczos algorithm.
Then $\Psi_{i_1 i_2 i_3 i_4}$ is used to construct the density matrix 
\begin{equation}
\rho_{i_1 i_2,i'_1 i'_2}=
\sum_{i_3 i_4}\Psi_{i_1 i_2 i_3 i_4}\Psi_{i'_1 i'_2 i_3 i_4}^*
\end{equation}
for the block containing the sites $n=1$ and $2$.
The density matrix specifies to what extent the basis states
$|i_1 \rangle|i_2 \rangle$ of the
block are contributing to the total wave function 
$|\Psi_{i_1 i_2 i_3 i_4} \rangle$.
This matrix is numerically diagonalized, and we obtain 
its eigenvalues $\lambda^\alpha$ and eigenvectors $v^\alpha_{i_1 i_2}$. 
Then we select the eigenvectors of the largest $m$ eigenvalues as new
basis states for the block. Here $m$ is the number of the basis states kept 
for the block at the next step.
Using the selected eigenvectors of the density matrix 
we represent the operators on the site, for example, $n=2$ as
\begin{equation}
(S_2)_{\alpha,\alpha'}=
\sum_{i_1 i_2 i'_2} (S_2)_{i_2,i'_2} (v^\alpha_{i_1 i_2})^* 
v^{\alpha'}_{i_1 i'_2}.
\end{equation}
A similar procedure is repeated for the block of the sites 
$n=3$ and $4$,
and all operators in the original system are 
represented in terms of the new basis states
\begin{eqnarray}
|\alpha\rangle&=&\sum_{i_1 i_2}v^\alpha_{i_1 i_2}|i_1\rangle |i_2\rangle 
, \\
|\beta\rangle&=&\sum_{i_3 i_4}v^\beta_{i_3 i_4}|i_3\rangle |i_4\rangle 
. 
\end{eqnarray}
To increase the size of system 
we introduce two new sites between 
the blocks $n=1,2$ and $n=3,4$. 
Using the basis states    
$|i_{2'}\rangle$ and $|i_{3'}\rangle$,
for the new sites, we construct the 
Hamiltonian matrix of the expanded system 
$H_{\alpha\, i_{2'} i_{3'} \beta, \alpha' \, i'_{2'} i'_{3'} \beta'}$. 
Renaming the indices as 
\begin{equation} 
\alpha\rightarrow i_1, i_{2'}\rightarrow i_2, 
i_{3'}\rightarrow i_3, \beta\rightarrow i_4, 
\label{DM_e}
\end{equation} 
we repeat the procedures from the diagonalization of the Hamiltonian
matrix.

The key feature of the above renormalization procedure is that 
the new basis states $|\alpha\rangle$ or $|\beta\rangle$ of each block 
contain the information that the block is a part of the total system.
As shown in Fig. \ref{DMRG1} the edge part of each block connecting 
to the remaining part 
of the system is located in the middle of the system, and this part 
is not so sensitive to the boundary conditions imposed on the 
total system. Thus we expect that the new basis states also 
dominantly contribute to 
the ground-state wave function of the expanded system which has two 
additional sites in the middle of the two blocks.

\begin{figure}
  \epsfxsize=85mm \epsffile{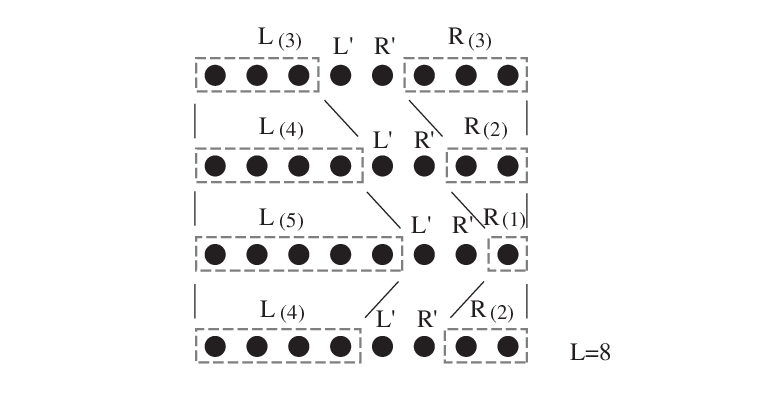}
\caption{
Schematic diagram of the finite system algorithm of the DMRG.
}
\label{DMRG2}
\end{figure}

The above algorithm is called the infinite system algorithm of the DMRG.
Using this algorithm we increase the size of the system.
In order to improve the basis states of the blocks,
it is necessary to fix the size of system and use the following 
algorithm which is known as the finite system algorithm 
of the DMRG. The schematic diagram of the finite system algorithm 
is shown in Fig. \ref{DMRG2}.

Let us take a block of size $n-1$ on the left and another block of 
size $n-1$ on the right whose 
basis states are represented by $|v_{L(n-1)}^i\rangle$ and 
$|v_{R(n-1)}^i\rangle$. 
These basis states and representations of the operators in the blocks are 
obtained after the ($n-2$)th renormalization step from the initial 
four-site system. 
The system of size $2n$ is constructed by 
inserting additional two sites for which the basis states are 
represented by $|i_{L'}\rangle$ and $|i_{R'}\rangle$.

We again diagonalize the Hamiltonian matrix of this $2n$-site system 
$H_{i_{L(n-1)} i_{L'} i_{R'} i_{R(n-1)}, i'_{L(n-1)} i'_{L'} 
i'_{R'} i'_{R(n-1)}}$
and obtain the ground-state wave function 
$\Psi_{i_{L(n-1)} i_{L'} i_{R'} i_{R(n-1)}}$.
Then construct the density matrix
\begin{eqnarray}
\lefteqn{\rho_{i_{L(n-1)} i_{L'},i'_{L(n-1)} i'_{L'}}=} \nonumber \\
& & \sum_{i_{R'} i_{R(n-1)}}\Psi_{i_{L(n-1)} i_{L'} i_{R'} i_{R(n-1)}}
 \Psi_{i'_{L(n-1)} i'_{L'} i_{R'} i_{R(n-1)}}^*
\end{eqnarray}
for the subspace spanned by $|v_{L(n-1)}^i\rangle$ and $|i_{L'}\rangle$.
We use the $m$ important eigenvectors of this density matrix
as a new basis states of the block containing $n$ sites,
and represent all operators in the new block in terms of the 
new basis states.

In the next step, we take the new left block with $n$ sites and 
the right block with $n-2$ sites.
The right block with $n-2$ sites has been obtained at the 
($n-3$)th renormalization step from the initial four-site system. 
Inserting two sites between these blocks,
we construct the Hamiltonian matrix of the total system 
$H_{i_{L(n)} i_{L'} i_{R'} i_{R(n-2)},  
i'_{L(n)} i'_{L'} i'_{R'} i'_{R(n-2)}}$, 
and repeat above all procedures. 

We continue to increase (decrease) the size of the left (right) block 
until the right 
block is reduced to the single site. Then we turn to decrease (increase) 
the size of the left (right) block in order to improve the basis states of the
right block. We continue to decrease (increase) the size of the left 
(right) block until the left block is reduced to the single site.
These procedures are continued back and forth until we get 
a good convergence.

In general, the total energy of the system is lowered as
the basis states of the blocks are reconstructed. 
Thus, the lowest energy is obtained after the convergence. 
The wave function obtained by the finite system algorithm of
the DMRG method may be represented by a matrix product form
\cite{Ostlund}.
Therefore  the finite system algorithm of the DMRG method 
is considered to be a numerical variational method which 
uses the matrix product wave function.  This is another reason why 
we can get remarkable accuracy by the DMRG method.
The accuracy of the ground-state energy and the wave function is 
determined by the eigenvalues of the density matrix which are
truncated out.  Thus we can improve the accuracy 
by increasing the number of basis states $m$ used in the 
calculations so long as the memory of computer allows.

\subsection{Finite temperature algorithm}

\begin{figure}
  \epsfxsize=85mm \epsffile{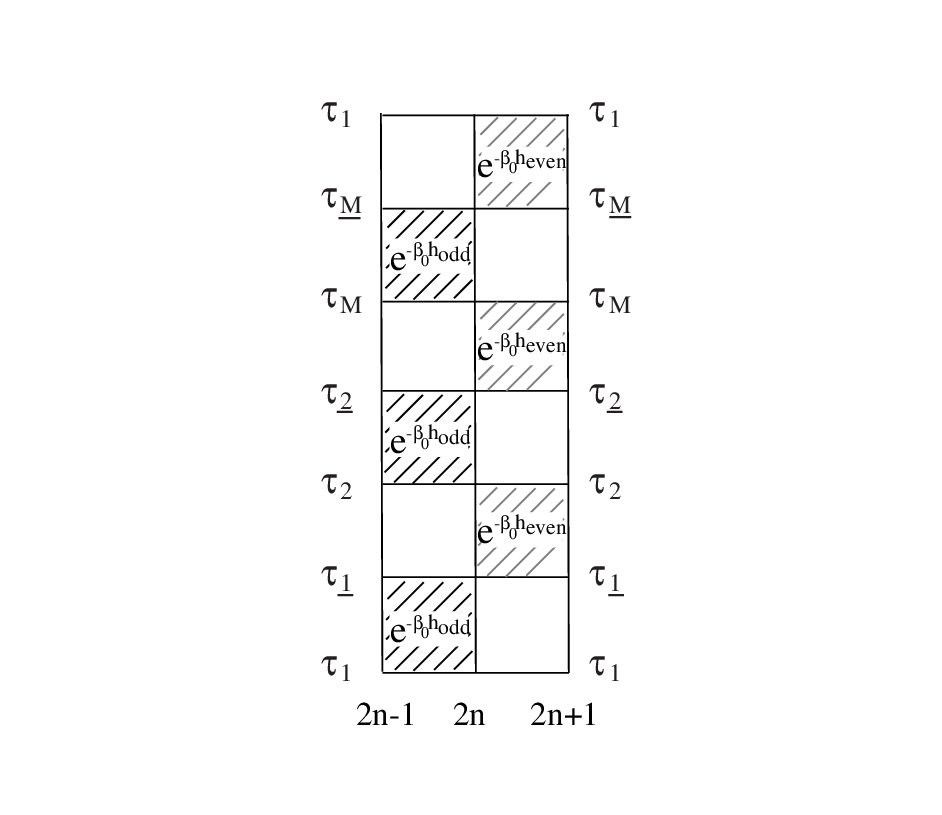}
\caption{
The quantum transfer matrix for $M=3$.
The $h_{\mbox{\scriptsize odd}}$ and $h_{\mbox{\scriptsize even}}$ 
represent $h_{2n-1,2n}$ and $h_{2n,2n+1}$, respectively.
$\Delta\tau=\tau_{i+1}-\tau_{i}=\beta_0$ and $\beta=M\beta_0$.
}
\label{TMfig}
\end{figure}

It is also possible to discuss thermodynamic quantities 
by the DMRG method, finite-$T$ DMRG.
The readers who are not interested in the detail of the iteration procedures 
may skip the paragraph including Eqs. (\ref{TM_rep}) to (\ref{TMtrans}).
In this method we use the quantum transfer matrix defined as
\begin{eqnarray}
{\cal T}_{n(M)}&=&[e^{-\beta h_{2n-1,2n}/M} e^{-\beta h_{2n,2n+1}/M} ]^M\nonumber \\ 
&=&\sum_{\sigma_{2n \tau_1}} \sum_{\sigma_{2n \tau_{\underline 1}}}\sum_{\sigma_{2n \tau_2}}\cdots 
\sum_{\sigma_{2n \tau_M}}\sum_{\sigma_{2n \tau_{\underline M}}}\nonumber\\
&&\langle \sigma_{2n-1, \tau_1} \sigma_{2n-1, \tau_{\underline 1}}|
e^{-\beta h_{2n-1,2n}/M} |\sigma_{2n, \tau_1} \sigma_{2n, \tau_{\underline 1}}\rangle \nonumber\\
&& \langle \sigma_{2n, \tau_{\underline 1}} \sigma_{2n, \tau_2}|
e^{-\beta h_{2n,2n+1}/M} |\sigma_{2n+1, \tau_{\underline 1}} \sigma_{2n+1, \tau_2}\rangle \nonumber\\
&& \langle \sigma_{2n-1, \tau_2} \sigma_{2n-1, \tau_{\underline 2}}|
e^{-\beta h_{2n-1,2n}/M} |\sigma_{2n, \tau_2} \sigma_{2n, \tau_{\underline 2}}\rangle \nonumber\\
&       & \cdots \nonumber \\
&& \langle \sigma_{2n, \tau_{\underline M}} \sigma_{2n, \tau_{1}}|
e^{-\beta h_{2n,2n+1}/M} |\sigma_{2n+1, \tau_{\underline M}} 
\sigma_{2n+1, \tau_{1}}\rangle. \nonumber\\
\label{TM}
\end{eqnarray}
This quantum transfer matrix is graphically shown in Fig. \ref{TMfig}.
Here $M$ is the Trotter number and 
$\tau_i$ is the discretized imaginary time whose 
intervals $\tau_{i+1}-\tau_i=\beta_0=\beta/M$.
In Eq. (\ref{TM}) $\sigma_{2n, \tau_{i}}$ represents states of the
site $2n$ corresponding to a given imaginary time $\tau_i$.
The Hamiltonian $H$ is assumed to be decomposed into two parts 
$H_{\mbox{\scriptsize odd}}=\sum_{n=1}^{L/2}h_{2n-1,2n}$ and 
$H_{\mbox{\scriptsize even}}=\sum_{n=1}^{L/2}h_{2n,2n+1}$
so that we can evaluate the matrix element of the
exponential function.  
Since the partition function $Z$ is given by the trace of the 
product of the quantum transfer matrix 
\begin{eqnarray}
Z & = & \mbox{Tr}\ e^{-\beta H} \nonumber \\ 
  & = & \lim_{M \rightarrow \infty}  \mbox{Tr}\  
         ( e^{-\beta H_{\mbox{\tiny odd}}/M} 
               e^{-\beta H_{\mbox{\tiny even}}/M} )^M \nonumber\\
  & = & \lim_{M \rightarrow \infty}  \mbox{Tr}\  
        [\prod_{n=1}^{N/2} ( e^{-\beta h_{2n-1,2n}/M} )
         \prod_{n=1}^{N/2} ( e^{-\beta h_{2n,2n+1}/M} ) ]^M \nonumber\\
 & = & \lim_{M\rightarrow \infty}\mbox{Tr}
       \left[\prod_{n=1}^{L/2}{\cal T}_n(M)\right],
\end{eqnarray}
thermodynamic properties of $L\rightarrow\infty$
are determined by the maximum eigenvalue and its eigenvectors.
To obtain the eigenvalue and the eigenvectors for  
a large Trotter number $M$, we iteratively increase the size of 
the quantum transfer matrix using a similar algorithm to 
the zero-temperature DMRG.

We first represent the quantum transfer matrix as
\begin{eqnarray}
{\cal T}_{(M)}
&=& \left\{
\begin{array}{ll}
{\cal T}^{A}_{(M)} 
{\cal T}^{B}_{(M)},
\hspace{0.3cm} \mbox{for $M$ : even} \\
{\cal T}^{A'}_{(M)} 
{\cal T}^{B'}_{(M)},
\hspace{0.3cm} \mbox{for $M$ : odd} 
\end{array}
\right. 
\label{TM_rep}
\end{eqnarray}
Thus the transfer matrix for $M=2$ is
\begin{eqnarray}
{\cal T}_{(M=2)}(\sigma_{2n-1,\tau_1} \sigma_{2n-1,\tau_{\underline 1}}
\sigma_{2n-1,\tau_2} \sigma_{2n-1,\tau_{\underline 2}}; \nonumber \\
\sigma_{2n+1,\tau_1} \sigma_{2n+1,\tau_{\underline 1} 
\sigma_{2n+1,\tau_2} \sigma_{2n+1,\tau_{\underline 2}})} \nonumber
\end{eqnarray}

\begin{figure}
  \epsfxsize=85mm \epsffile{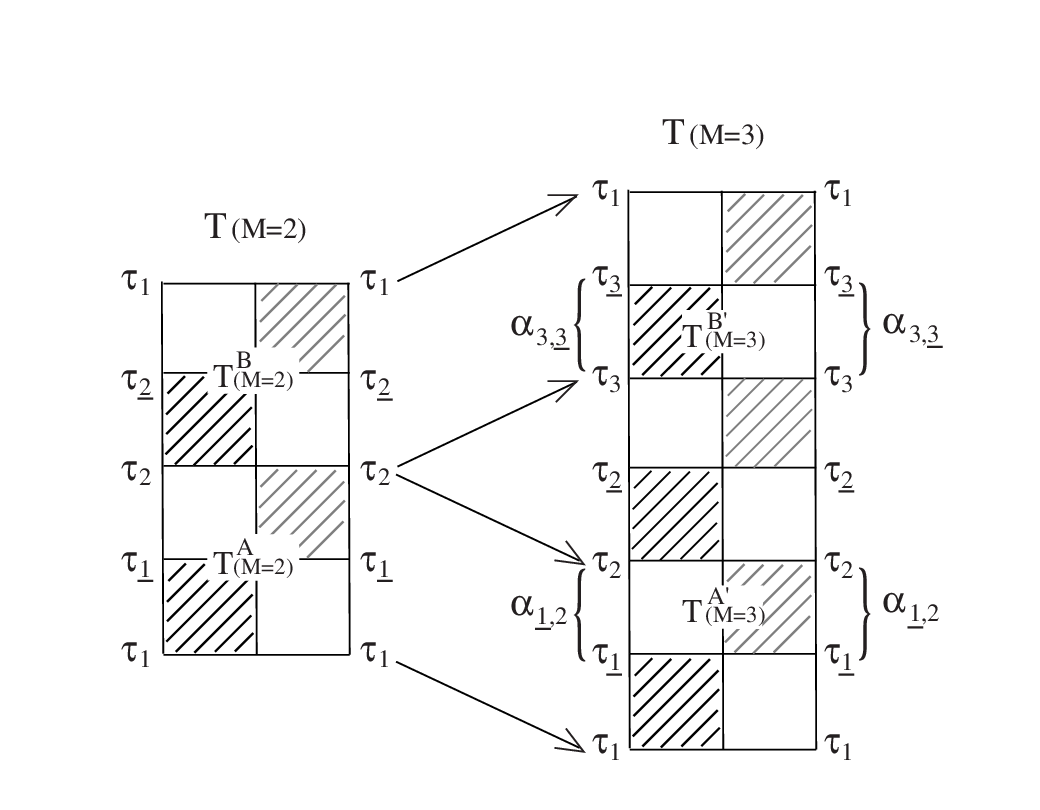}
\caption{
Increasing of $M$ of the quantum transfer natrix ${\cal T}$.
The $\alpha_{n,m}$ is the indices of the new basis states.
}
\label{TM_exp}
\end{figure}

\vspace{-0.6cm}
\begin{eqnarray}
&=& 
\sum_{\sigma_{2n,\tau_1}}
\sum_{\sigma_{2n,\tau_2}}
\nonumber \\
&& {\cal T}^A_{(M=2)}(\sigma_{2n-1,\tau_1} \sigma_{2n-1,\tau_{\underline 1}} \sigma_{2n,\tau_2};
\sigma_{2n,\tau_1} \sigma_{2n+1,\tau_{\underline 1}} \sigma_{2n+1,\tau_2}) \nonumber \\
&& {\cal T}^B_{(M=2)}( \sigma_{2n-1,\tau_2}  \sigma_{2n-1,\tau_{\underline 2}} \sigma_{2n,\tau_1};
\sigma_{2n,\tau_2} \sigma_{2n+1,\tau_{\underline 2}} \sigma_{2n+1,\tau_1}).\nonumber \\ 
\end{eqnarray}
\noindent
Here we have introduced ${\cal T}^A_{(M=2)}$ and $ {\cal T}^B_{(M=2)}$
which are defined as
\begin{eqnarray}
\lefteqn{{\cal T}^A_{(M=2)}(\sigma_{2n-1,\tau_1} \sigma_{2n-1,\tau_{\underline 1}} \sigma_{2n,\tau_2};
\sigma_{2n,\tau_1} \sigma_{2n+1,\tau_{\underline 1}} \sigma_{2n+1,\tau_2})} \nonumber \\
&=& \sum_{\sigma_{2n,\tau_{\underline 1}}} 
\langle \sigma_{2n-1, \tau_{1}} \sigma_{2n-1, \tau_{\underline 1}}|
e^{-\beta h_{odd}/M} |\sigma_{2n, \tau_{1}} 
\sigma_{2n, \tau_{\underline 1}}\rangle  \nonumber \\
&& \langle \sigma_{2n, \tau_{\underline 1}} \sigma_{2n, \tau_2}|
e^{-\beta h_{even}/M} |\sigma_{2n+1, \tau_{\underline 1}} \sigma_{2n+1, \tau_2}\rangle ,
\end{eqnarray}
\begin{eqnarray}
\lefteqn{{\cal T}^B_{(M=2)}(\sigma_{2n-1,\tau_2} \sigma_{2n-1,\tau_{\underline 2}} \sigma_{2n,\tau_{1}};
\sigma_{2n,\tau_2} \sigma_{2n+1,\tau_{\underline 2}} \sigma_{2n+1,\tau_{1}})} \nonumber \\
&=& \sum_{\sigma_{2n,\tau_{\underline 2}}}
\langle \sigma_{2n-1, \tau_2} \sigma_{2n-1, \tau_{\underline 2}}|
e^{-\beta h_{odd}/M} |\sigma_{2n, \tau_2} \sigma_{2n, \tau_{\underline 2}}\rangle  \nonumber \\
&&\langle \sigma_{2n, \tau_{\underline 2}} \sigma_{2n, \tau_{1}}|
e^{-\beta h_{even}/M} |\sigma_{2n+1, \tau_{\underline 2}} 
\sigma_{2n+1, \tau_{1}}\rangle .
\end{eqnarray}
where
$h_{\mbox{\scriptsize odd}}=h_{2n-1,2n}$ and  
$h_{\mbox{\scriptsize even}}=h_{2n,2n+1}$.
Then we iteratively increase $M$ of ${\cal T}^A_{(M)}$ and $ {\cal T}^B_{(M)}$ as
\begin{eqnarray}
{\cal T}^A_{(M)} e^{-\beta_0 h_{\mbox{\tiny odd}}} & \rightarrow & 
{\cal T}^{A'}_{(M+1)} \label{TM_A}\\
 e^{-\beta_0 h_{\mbox{\tiny even}}} {\cal T}^B_{(M)} & \rightarrow & 
{\cal T}^{B'}_{(M+1)},\label{TM_B}\\
{\cal T}^{A'}_{(M)} e^{-\beta_0 h_{\mbox{\tiny even}}} & \rightarrow & 
{\cal T}^{A}_{(M+1)} \label{TM_A'}\\ 
 e^{-\beta_0 h_{\mbox{\tiny odd}}} {\cal T}^{B'}_{(M)} & \rightarrow & 
{\cal T}^{B}_{(M+1)}.
\end{eqnarray}
The example for the increase of $M$, Eqs. (\ref{TM_A}) and (\ref{TM_B}), 
for $M=2$ is graphically shown in Fig. \ref{TM_exp}.

In order to represent the transfer matrix in a restricted number of 
basis states, 
we have to select important basis states which have significant 
weight for the representation of the transfer matrix. 
For this purpose we use the generalized asymmetric density matrix
similar to that used in the zero-temperature DMRG.
For example, the density matrix which we use in the 
procedure Eq. (\ref{TM_A}) for $M=2$ is
\begin{eqnarray}
\lefteqn{\rho(\sigma_{2n-1,\tau_{\underline 1}} \sigma_{2n-1,\tau_2};
\sigma_{2n+1,\tau_{\underline 1}} \sigma_{2n+1,\tau_2})} \nonumber \\
&=& \sum_{\sigma_{\tau_1}}\sum_{\sigma_{\tau_{\underline 2}}}
V^L(\sigma_{\tau_1} \sigma_{2n-1,\tau_{\underline 1}},
\sigma_{2n-1,\tau_2} \sigma_{\tau_{\underline 2}}j\nonumber \\
& & V^Ri\sigma_{\tau_1} \sigma_{2n+1,\tau_{\underline 1}},
\sigma_{2n+1,\tau_2} \sigma_{\tau_{\underline 2}}) 
\end{eqnarray}
where $V^L$ and $V^R$ are the left and right eigenvectors of ${\cal T}_{(M=2)}$
which have the maximum eigenvalue.
The $V^L$ and $V^R$ are 
generally different owing to the non-Hermite property
of the transfer matrix. 
The diagonalization of the density matrix provides eigenvectors,
$v^L_\alpha$ and $v^R_\alpha$,
which satisfies the equations
\begin{eqnarray}
 \sum_{\sigma_{\tau_{\underline 1}} \sigma_{\tau_2}}v^L_\alpha(\sigma_{\tau_{\underline 1}} \sigma_{\tau_2})
 \rho(\sigma_{\tau_{\underline 1}} \sigma_{\tau_2};\sigma'_{\tau_{\underline 1}} \sigma'_{\tau_2})
 & = & \gamma_\alpha 
v^L_\alpha(\sigma'_{\tau_{\underline 1}} \sigma'_{\tau_2}) \\
 \sum_{\sigma_{\tau_{\underline 1}} \sigma_{\tau_2}}
 \rho(\sigma'_{\tau_{\underline 1}} \sigma'_{\tau_2};
\sigma_{\tau_{\underline 1}} \sigma_{\tau_2})
v^R_\alpha(\sigma_{\tau_{\underline 1}} \sigma_{\tau_2})
 & = & \gamma_\alpha v^R_\alpha(\sigma'_{\tau_{\underline 1}} \sigma'_{\tau_2}) .
\end{eqnarray}
We select the $m$ eigenvectors which have the largest eigenvalues $\gamma_\alpha$,
and we use them as the new basis states
\begin{eqnarray}
\langle \alpha_{2n-1}| & = & \sum_{\sigma_{2n-1,\tau_{\underline 1}}} \sum_{\sigma_{2n-1,\tau_2}} \nonumber \\
& & v^L_\alpha(\sigma_{2n-1,\tau_{\underline 1}} \sigma_{2n-1,\tau_2})
\langle \sigma_{2n-1,\tau_{\underline 1}} \sigma_{2n-1,\tau_2}| ,\\
| \alpha_{2n+1} \rangle
& = & 
\sum_{\sigma_{2n+1,\tau_{\underline 1}}} \sum_{\sigma_{2n+1,\tau_2}} \nonumber\\
& & v^R_{\alpha}(\sigma_{2n+1,\tau_{\underline 1}} \sigma_{2n+1,\tau_2})
|\sigma_{2n+1,\tau_{\underline 1}} \sigma_{2n+1,\tau_2}\rangle.
\end{eqnarray}
Then we represent ${\cal T}^{A'}_{(M=3)}$ as
\begin{eqnarray}
{\cal T}^{A'}_{(M=3)} 
(\sigma_{2n-1,\tau_1} \alpha_{2n-1,\tau_{{\underline 1},2}} \sigma_{2n-1,\tau_{\underline 2}};
\sigma_{2n,\tau_1} \alpha'_{2n+1,\tau_{{\underline 1},2}} \sigma_{2n,\tau_{\underline 2}}) \nonumber
\end{eqnarray}

\vspace{-0.5cm}
\begin{eqnarray}
& = & \sum_{\sigma_{2n-1,\tau_{\underline 1}}} \sum_{\sigma_{2n-1,\tau_2}}
\sum_{\sigma_{2n+1,\tau_{\underline 1}}} \sum_{\sigma_{2n+1,\tau_2}}
\sum_{\sigma_{2n,\tau_2}} \nonumber \\
& & v^R_\alpha(\sigma_{2n-1,\tau_{\underline 1}} \sigma_{2n-1,\tau_2})
 v^L_{\alpha'}(\sigma_{2n+1,\tau_{\underline 1}} \sigma_{2n+1,\tau_2}) \nonumber \\
& & {\cal T}^{A}_{(M=2)}(\sigma_{2n-1,\tau_1} \sigma_{2n-1,\tau_{\underline 1}} \sigma_{2n,\tau_2};
\sigma_{2n,\tau_1} \sigma_{2n+1,\tau_{\underline 1}} \sigma_{2n+1,\tau_2}) \nonumber \\
& & \langle \sigma_{2n-1, \tau_{2}} \sigma_{2n-1, \tau_{\underline 2}}|
e^{-\beta h_{odd}/M} |\sigma_{2n, \tau_{2}} 
\sigma_{2n, \tau_{\underline 2}}\rangle .
\label{TMtrans}
\end{eqnarray}
We repeat similar procedure for Eqs. (\ref{TM_A}) to (\ref{TMtrans}) 
and obtain the maximun eigenvalue 
and its eigenvectros of the quantum transfer matrix for a desired $M$.

Compared with the zero-temperature DMRG algorithm
the finite-$T$ DMRG is more subtle from the point of view of
numerical stability.
The asymmetric density matrix sometimes
yields complex eigenvalues, although in principle they must 
be real. To avoid those unphysical complex eigenvalues we 
need accurate numerical calculations taking account of various
symmetries of the system. 

The free energy per site of the infinite system is obtained from the 
maximum eigenvalue $\lambda$ of the transfer matrix as
$F=-(T/2) \ln \lambda $. 
Static quantities like specific heat and susceptibilities are 
obtained from the free energy.
The specific heat is calculated by the numerical derivatives of 
$F$ with respect to the temperature $T$.
The spin and charge susceptibilities are calculated by a shift of
$F$ under an applied magnetic field or chemical potential.

\begin{figure}
  \epsfxsize=85mm \epsffile{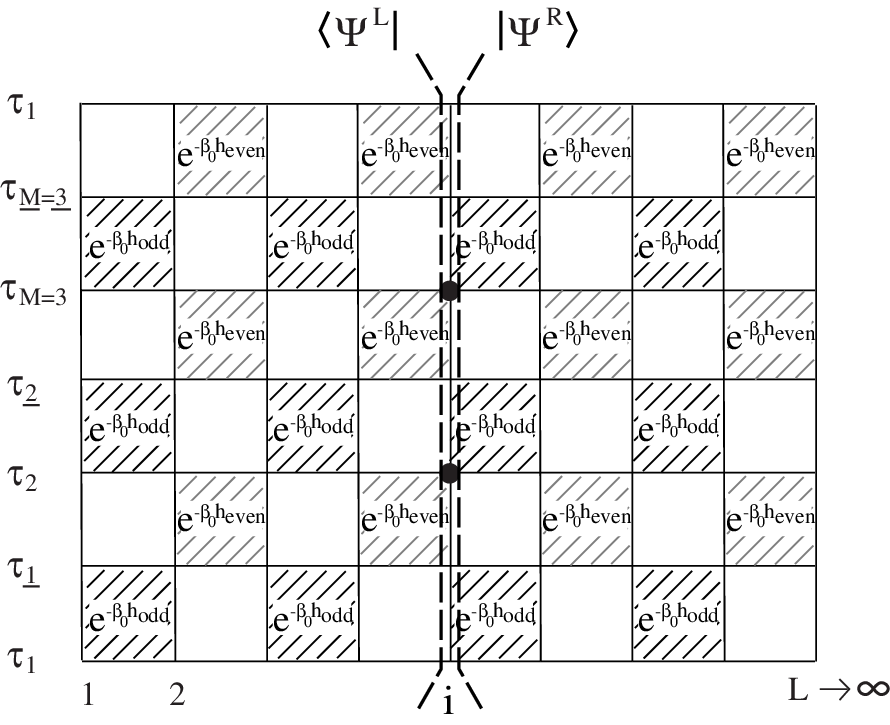}
\caption{
Schematic diagram of the calculation of the imaginary time
correlation function 
$\langle \Psi^L| c_{i \sigma}(\tau_3)\ c^\dagger_{i \sigma}(\tau_2) 
|\Psi^R\rangle$.
The Trotter number of this example is $M=3$.
}
\label{G_calc}
\end{figure}

The calculation of the dynamic quantities requires additional steps.
We first calculate correlation functions in the $\beta$ direction.
This calculation requires good accuracy for the eigenvectors of 
the transfer matrix $\langle \Psi^L|$ and $|\Psi^R\rangle$.
Thus it is necessary to use the finite system algorithm of the DMRG.
The Green's function in the $\beta$ direction is obtained
from the left and the right eigenvectors of the transfer matrix 
whose eigenvalue is the largest (See Fig. \ref{G_calc}): 
\begin{eqnarray}
G(\tau_j)
&\equiv& -\mbox{Tr}\{\mbox{e}^{-\beta H} c_{i \sigma}(\tau_j)\ 
c^\dagger_{i \sigma}(0)\}/Z \nonumber \\
&=& -\langle \Psi^L| c_{i \sigma}(\tau_j)\ c^\dagger_{i \sigma}(0) 
|\Psi^R\rangle . 
\label{G_tau}
\end{eqnarray}
Similarly a local dynamic correlation function $\chi_{AB}(\tau_j)$
is obtained as
\begin{eqnarray}
\chi_{AB}(\tau_j)
&\equiv& \mbox{Tr}\{\mbox{e}^{-\beta H} A_i(\tau_j)\ B_i(0)\}
/Z \nonumber \\
&=& \langle \Psi^L|A_i(\tau_j)\ B_i(0)|\Psi^R\rangle .
\end{eqnarray}
By Fourier transformation, 
the Green's function and the the dynamic correlation function 
as functions of the imaginary frequencies are obtined as
\begin{eqnarray}
G(i \omega_n) & = & \frac{\beta}{M} 
\sum_j \mbox{e}^{i \omega_n \tau_j} G(\tau_j), \\
\chi_{AB}(i \omega_n) & = & \frac{\beta}{M} 
\sum_j \mbox{e}^{i \omega_n \tau_j} \chi_{AB}(\tau_j),
\end{eqnarray} 
where $\omega_n$ is the Matsubara frequency that is
$\pi(2n+1)/\beta$ for fermionic operators and $2\pi n/\beta$ 
for bosonic operators.

The real frequency Green's function and dynamic susceptibility
are obtained by the analytic continuation to the real frequency axis.
We can use the Pad\'e approximations or the maximum entropy method
for this purpose.  The former method is based on the fittings of 
$G(i \omega_n)$ or $\chi_{AB}(i \omega_n)$ by rational funcitons
of frequency $i \omega_n$ which are analytically continued to 
the real axis by $i \omega_n \rightarrow \omega + i \delta$.

The maximum entropy method is based on the spectral representations
\begin{eqnarray}
G(\tau) &=& \int^{\infty}_{-\infty} 
  \rho(\omega) \frac{\mbox{e}^{- \tau \omega}}{1+\mbox{e}^{- \beta \omega} 
} d\omega , \\
\chi_{AB}(\tau) &=& \int^{\infty}_{-\infty} 
\frac{1}{\pi}
\mbox{Im}\chi_{AB}(\omega) 
\frac{\mbox{e}^{- \tau \omega}}{1-\mbox{e}^{- \beta \omega}} d\omega ,
\end{eqnarray} 
with $\rho(\omega)=-\frac{\mbox{Im}}{\pi}G(\omega+i\delta)$ 
being the density of state.
Starting from a flat spectrum, this method finally finds
optimal $\rho(\omega)$ and $\chi_{AB}(\omega)$ that reproduce 
$G(\tau)$ and $\chi_{AB}(\tau)$ best.

The dynamical structure factor $S_{AB}(\omega)$ is related to 
the imaginary part of $\chi_{AB}(\omega)$ through the fluctuation 
dissipation theorem,
\begin{equation}
\mbox{Im} \chi_{AB}(\omega) =  
  \pi (1 - \mbox{e}^{-\beta \omega} ) S_{AB}(\omega).
\label{S_omega}
\end{equation} 

In section III we use the standard zero-temperature DMRG
method to study the ground-state properties of the paramagnetic 
metallic phase.  In section IV after a brief discussion of the 
ground-state properties by using the zero-temperature DMRG, 
finite temperature properties of the Kondo spin liquid phase 
will be discussed extensively by using the finite-$T$ DMRG.

\section{TOMONAGA-LUTTINGER LIQUID PROPERTIES OF THE PARAMAGNETIC
METALLIC PHASE}

This section concerns with the paramagnetic phase away from 
half-filling, see Fig.\ref{phase-dgm}. Since there is no symmetry 
breaking it is natural to consider that away from half-filling 
the translationally invariant Kondo lattice model is metallic.  
In one dimension it is well known that various interacting 
metallic systems including the Hubbard model and the $t$-$J$ 
model belong to the universality class of 
Tomonaga-Luttinger liquids\cite{Haldane}. Therefore, the 
first question concerning the paramagnetic metallic phase 
of the Kondo lattice model is whether it belongs to this class 
or not.

The spin-1/2 Tomonaga-Luttinger liquids have gapless charge and 
spin excitations.  In one dimension the charge excitations are 
characterized by the velocity of the charge density $v_\rho$ and 
the correlation exponemt $K_\rho$.  Similarly, the spin excitations 
are characterized by the velocity of the spin density $v_\sigma$, 
but the correlation exponent in the spin sector is fixed by the 
SU(2) symmetry, $K_\sigma =1 $.
The low-energy physics of a Tomonaga-Luttinger liquid is 
completely determined when these parameters are obtained.
For example, the spin and charge susceptibilities are given by
\begin{eqnarray}
  \chi_\sigma &=& \frac{2}{\pi v_\sigma}\ , 
\label{schi}\\
  \chi_\rho   &=& \frac{2K_\rho}{\pi v_\rho}\ ,
\label{cchi}
\end{eqnarray}

Reflecting the gapless excitations, the density-density 
 and spin-spin correlation functions show power-law 
decays where the exponents are determined by the correlation 
exponent, $K_\rho$.  
The asymptotic forms of the density-density and spin-spin 
correlation functions are
\begin{eqnarray}
\langle n(x)n(0)\rangle&=&K_\rho/(\pi x)^2+A_1\cos(2k_Fx)
x^{-(1+K_\rho)}  \nonumber \\
 & & + A_2\cos(4k_Fx)x^{-4K_\rho}\ , 
\label{ddcorr}\\
\langle S(x)\cdot S(0)\rangle&=&
1/(\pi x)^2+B_1\cos(2k_Fx)x^{-(1+K_\rho)}\ ,
\label{sscorr}
\end{eqnarray}
where $k_F=\pi\rho/2$, with $\rho$ being the density of charge
carriers, is the Fermi momentum.  The logarithmic corrections 
are neglected in Eqs.(\ref{ddcorr}) and (\ref{sscorr})\cite{Schulz}.
 
For the Hubbard model or the $t$-$J$ model, the definition of the 
density of carriers is straighforward.  On the other hand, for the Kondo 
lattice model it is already questionable.  When we take naively the 
conduction electrons as carriers, then the Fermi momentum is given 
by $k_F=k_{Fs}=\pi n_c/2$.  However,
a different point of view is possible.  Let us consider the Kondo
lattice model as an effective Hamiltonian for the periodic Anderson
model.  For the latter, the density of carriers is the sum of  the $f$
electon density and the conduction electron density.  According to 
the Luttinger sum rule \cite{Luttinger} the position of Fermi points 
do not change when the interaction is increased as long as the ground 
state remains paramagnetic.  Therefore this property may be
carried over to the Kondo lattice model and it would be also natural 
to assume $k_F=k_{Fl}=\pi(1+n_c)/2$ for the Kondo lattice model.

Concerning the paramagnetic metallic phase, there are two basic 
questions:\\
(1) Is it a Tomonaga-Luttinger liquid?\\
(2) If it is the case, what is the size of the Fermi momentum?
Is it large, $k_{Fl}=\pi(1+n_c)/2$ or small, $k_{Fs}=\pi n_c/2$?
The DMRG is a powerful method to address these questions. 

Let us define the ground-state energy in a given spin-$S$ subspace 
for a finite system with $L$-sites by $E_g(L, N_c, S)$ where $N_c$ 
is the number of conduction electrons. In the 
following we will consider only even $L$ and $N_c$ and thereby 
integer $S$. The spin gap of a finite system is defined by 
\begin{equation}
   \Delta_s(L)=E_g(L, N_c, S=1)-E_g(L, N_c, S=0)\ .
\label{spin_gap_def}
\end{equation}
Concerning the charge excitations we study the difference of the
chemical potentials, $\mu_+ - \mu_-$ which is given by
\begin{eqnarray}
   2\mu_+ &=& E_g(L, N_c+2, S=0) - E_g(L, N_c, S=0)\ , \\
   2\mu_- &=& E_g(L, N_c-2, S=0) - E_g(L, N_c, S=0)\ .
\end{eqnarray}

\begin{figure}
  \epsfxsize=85mm \epsffile{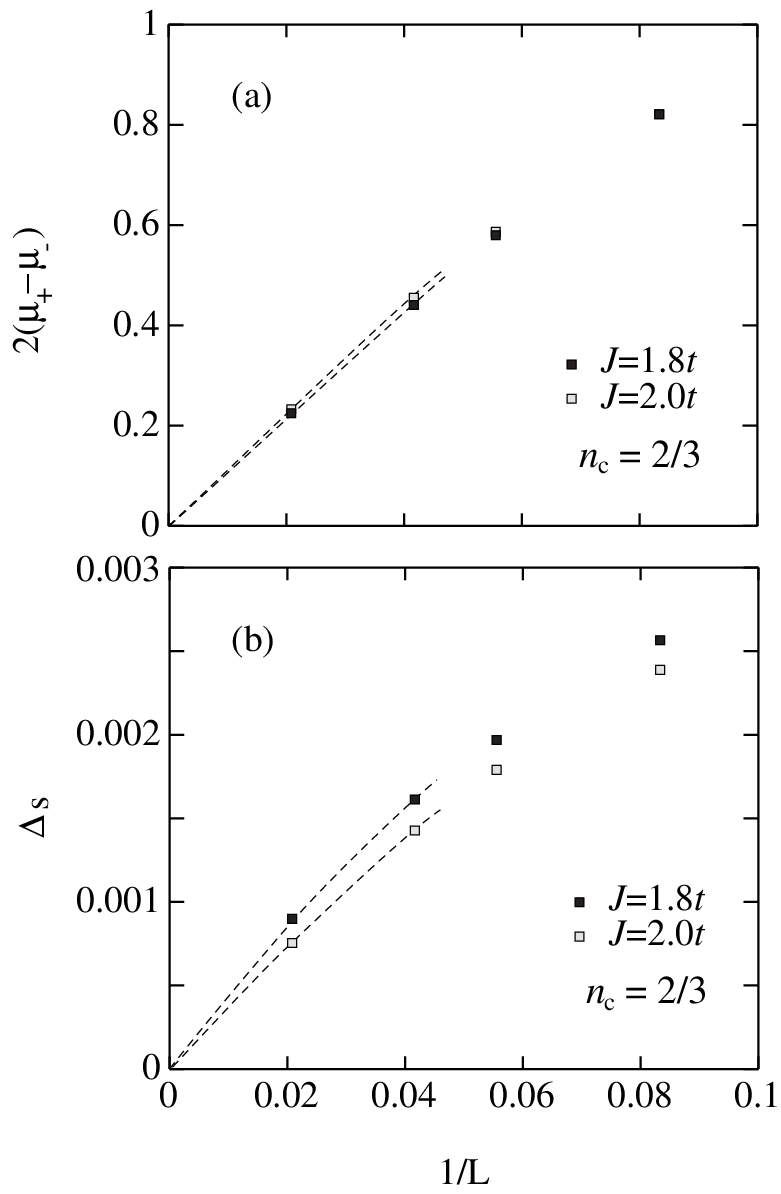}
\caption{
(a) Size dependence of the difference of the chemical
potentials, $\mu_+-\mu_-$, in the one dimensional Kondo lattice
model.
(b) Size dependence of the spin gap.
$L$ is the system size and the density of conduction 
electrons is fixed to $n_c=2/3$. The energy unit is $t$.
Typical truncation errors in the DMRG calculations are $10^{-4}$.
}
\label{ChargeGap}
\end{figure}

Figure \ref{ChargeGap} shows (a) the spin excitation gap and 
(b) the difference of
the chemical potentials as a function of inverse of the system size.
For the example the density of conduction electrons are fixed to 
$n_c =2/3$\cite{STU}.
Both quatities go to zero as $1/L \rightarrow 0$, which means that 
the spin excitations and the charge excitations are gapless.  
Therefore it is most likely that the paramagnetic metallic phase of
the Kondo lattice model belongs to the universality class of the 
Tomonaga-Luttinger liquids. 

Now we will determine the parameters of the Tomonaga-Luttinger liquid.
From the slope of the spin gap the velocity of the spin excitations
are determined by 
\begin{equation}
    \Delta_s(L)=v_\sigma\pi/L,
\end{equation} 
since the lowest spin excitation of a finite system with the open boundary
condition has the momentum $\pi/L$.  
The difference of the chemical potentials is related with the charge 
susceptibility by 
\begin{equation}
   \mu_+ - \mu_- = \frac{2}{\chi_\rho L}\ .
\end{equation}
The values in Table I for $v_\sigma$ and $\chi_\rho$ are determined
from the slopes.  The values of the spin susceptibility in the Table I
is obtained from the spin velocity through Eq.(\ref{schi}). 

To determine the charge velocity and the correlation exponent
separately, another independent measurement is necessary.
The correlation exponent $K_\rho$ may be determined from 
the density-density or the spin-spin correlation functions.  
However, determination of the exponent of a power law decay is the most 
difficult for any numerical calculations.  In order to determine 
$K_\rho$ we need to see the long-range behaviors of the correlation
functions with sufficient accuracy. Instead of looking at the correlation
functions we looked at the Friedel oscillations because the latter are 
numerically more reliable than the former\cite{STU,SUNI}.  
The reason for this is 
that the correlation functions are site off-diagonal, while the 
Friedel oscillations are site diagonal.

\begin{table}
\caption{
Luttinger liquid parameters of the one dimensional Kondo
lattice model. The carrier density $n_c$ is $2/3$.  The energy
unit is $t$. The errors are estimated from the ambiguity of the
power law decay of the charge density Friedel oscillations.
}
\begin{tabular}{lccccc}
   & $K_\rho$ & $v_\sigma$ & $\chi_\sigma$ &
                         $v_\rho$ & $\chi_\rho$ \\ \hline
$J/t=0   $ & 1    & -     & -    & 1.73 & 0.37 \\
$J/t=1.5$ & 0.19 $\pm$ 0.03 &       &    & 0.30 $\pm$ 0.06 & 0.42 \\
$J/t=1.8$ & 0.24 $\pm$ 0.02 & 0.014 & 46 & 0.41 $\pm$ 0.06 & 0.38 \\
$J/t=2.0$ & 0.27 $\pm$ 0.02 & 0.011 & 56 & 0.48 $\pm$ 0.06 & 0.36 \\
\end{tabular}
\label{table-1}
\end{table}


The Friedel oscillations are density oscillations
induced by a local perturbation.
In a Tomonaga-Luttinger liquid, power law anomalies in
correlation functions naturally reflect themselves
in the Friedel oscillations.  
The usual Friedel oscillations induced by an impurity potential 
are given by
\begin{eqnarray}
\delta \rho(x)& \sim & C_1\cos(2k_Fx)x^{(-1-K_\rho)/2}
 + C_2\cos(4k_Fx)x^{-2K_\rho}
\nonumber \\
\label{CDO}
\end{eqnarray}
as a function of the distance $x$ from the impurity
\cite{wire1,wire2,wire3}. 
Analogously, spin density oscillations
induced by a local magnetic field behave as
\begin{eqnarray}
 \sigma(x) & \sim & D_1\cos(2k_Fx)x^{-K_\rho}.
\label{SDO}
\end{eqnarray}
Thus, we can determine $K_\rho$ from
the asymptotic form of the oscillations.
It is worth to be noted that origin of the RKKY interaction 
may be traced back to the spin density oscillations induced by 
a localized spin.

\begin{figure}
  \epsfxsize=85mm \epsffile{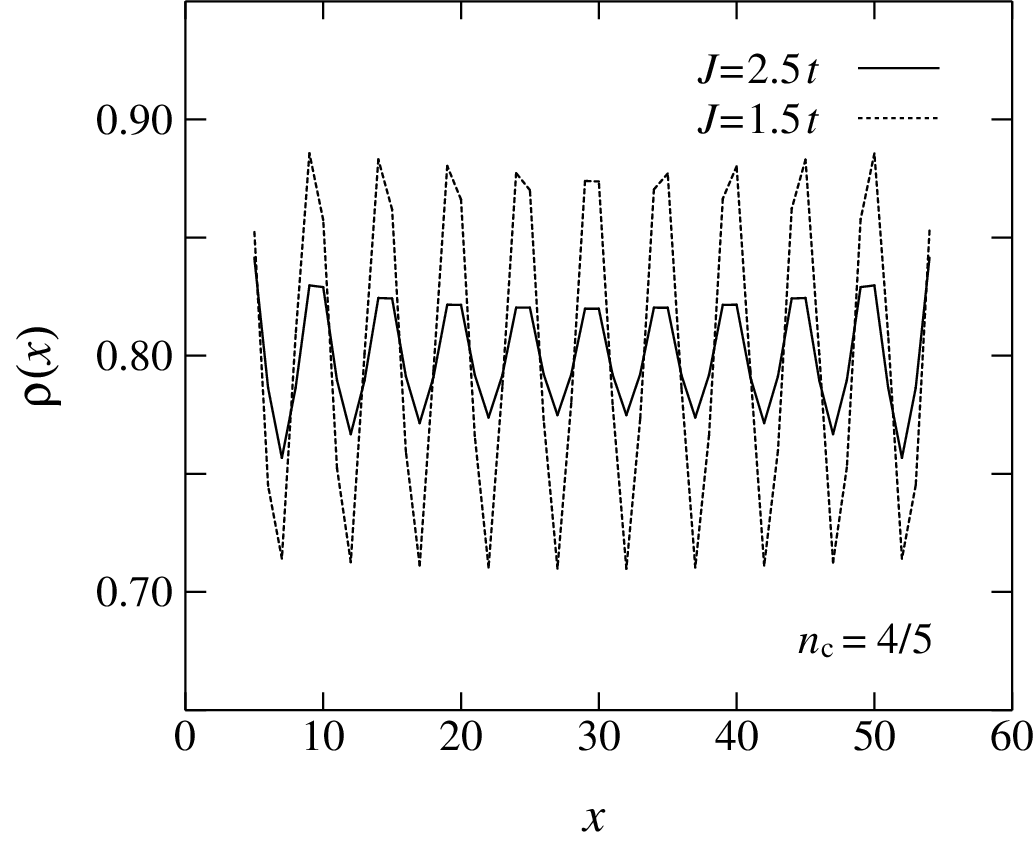}
\caption{
Charge density oscillations of the Kondo lattice model.
The system size is 60 sites and the carrier density is $n_c=4/5$.
The solid line and the broken line correspond to $J=2.5t$
and $J=1.5t$, respectively.
Typical truncation errors in the DMRG calculations
are $1 \times 10^{-6}$ for $J=2.5t$
and $3 \times 10^{-6}$ for $J=1.5t$.
}
\label{charge-r}
\end{figure}

\begin{figure}
  \epsfxsize=85mm \epsffile{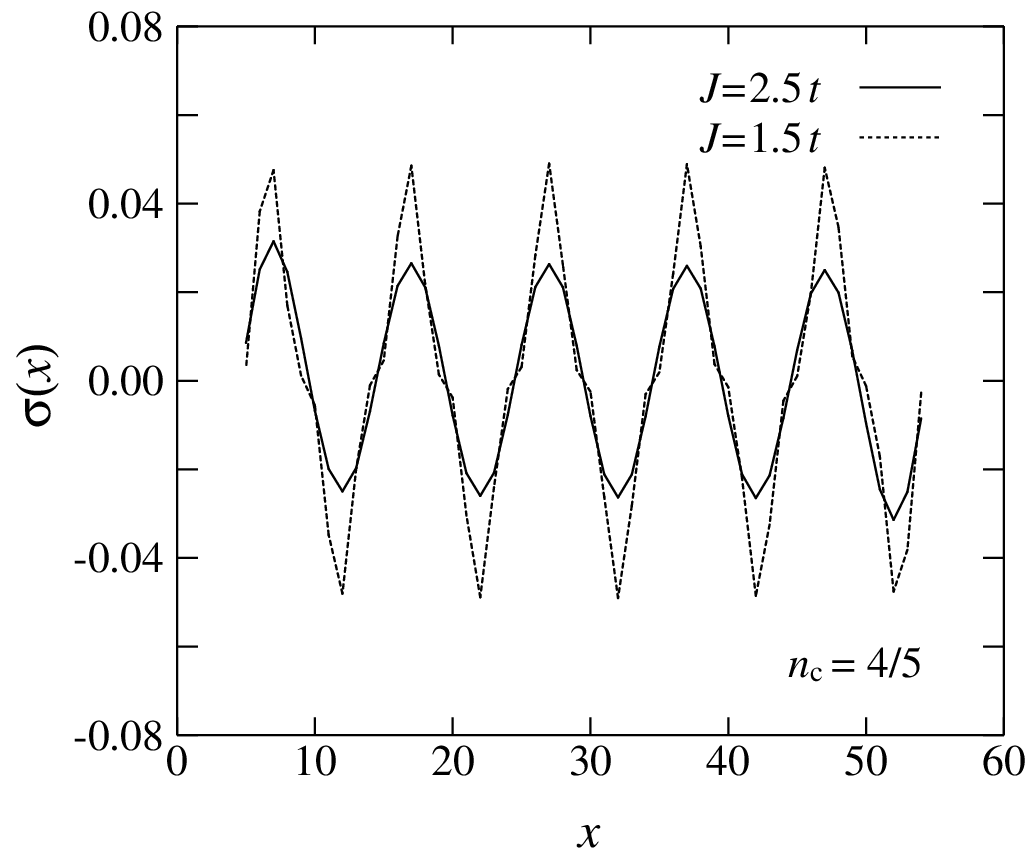}
\caption{
Spin density oscillations of the Kondo lattice model.
The system size is 60 sites and the carrier density is $n_c=4/5$.
The solid line and the broken line correspond to $J=2.5t$
and $J=1.5t$, respectively. The strength of the local magnetic
field $h$ is $0.1t$.
Typical truncation errors in the DMRG calculations
are $1 \times 10^{-6}$ for $J=2.5t$
and $3 \times 10^{-6}$ for $J=1.5t$.
}
\label{spin-r}
\end{figure}

The charge density oscillations induced by the open boundary 
conditions are shown in Fig.\ref{charge-r} for $J=1.5t$ and $J=2.5t$ at 
the density $n_c=4/5$\cite{SUNI}.  The spin density oscillations 
induced by the local magnetic fields applied at the both ends with 
opposite directions are shown in Fig.\ref{spin-r} for the same set 
of parameters\cite{SUNI}.  It is clearly seen that the dominant period of
the charge density oscillations is five sites, $q=2\pi/5$, while 
for the spin density oscillations ten sites, $q=\pi/5$.    

In the strong coupling limit of the Kondo lattice model, each conduction
electron form a local singlet with the $f$ spin on the same site.  
However away from half-filling these singlets can move in the 
lattice with the effective 
hopping matrix elements reduced by half.  Equivalently, we can regard 
the unpaired $f$ spins as mobile objects with the reduced hopping
energy $t/2$ with its sign reversed.  Note that in the original model
hopping matrix elements are defined by $-t$. 
Thus the effective model of the strong coupling limit
is the $t$-model where the number of carriers is $L-N_c$ and double 
occupancy of the carriers is prohibited. 

The charge response of the system is identical to the spinless
Fermions where the Fermi point is given by $\pi-\pi(1-n_c)=\pi
n_c$.  Therefore the induced charge density shows oscillations
corresponding to $2\pi n_c$ which is equivalent to $4k_{Fs}$ and $4k_{Fl}$.
This analysis shows that in the strong coupling limit the amplitude of
the $4k_F$ oscillations dominates over the $2k_F$ oscillations
for the charge Friedel oscillations, Eq.(\ref{CDO}).  The period of five sites
is naturally understood in this way and we have confirmed that 
for a weaker coupling the amplitude of the $2k_F$ oscillation develops.

\begin{figure}
  \epsfxsize=85mm \epsffile{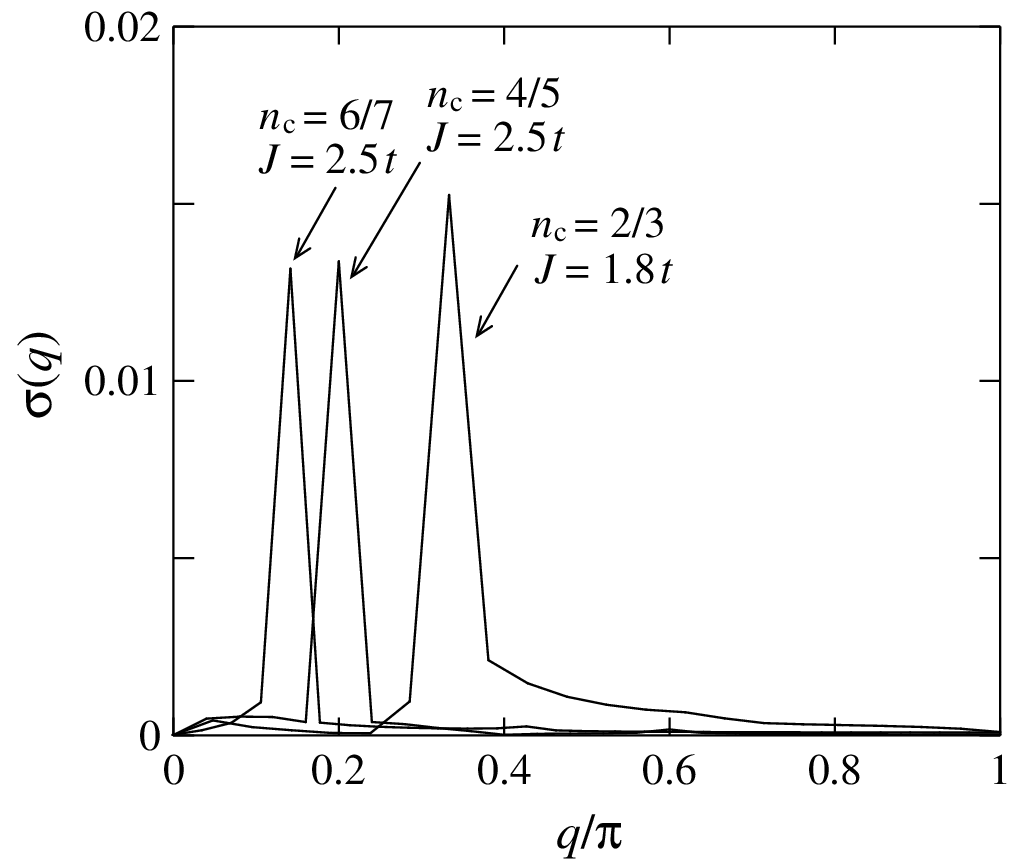}
\caption{
Fourier components of the spin density Friedel oscillations.
}
\label{FSFO}
\end{figure}

When we consider the spins of the unpaired $f$ electrons, there 
remains a macroscopic $2^{L-N_c}$-fold degeneracy.  
Concerning the spin 
sector the lifting of the degeneracy is essential. For the specific 
model where the degeneracy is lifted by the next nearest neighbor
hoppings it is shown analytically that the Fermi surface is big, 
$k_F=k_{Fl}$\cite{UNT}.  For a finite $J$ the degeneracy of the 
Kondo lattice model is always lifted.
The period of ten sites of the spin density Friedel oscillations 
shown in Fig. \ref{spin-r} indicate that the 
$2k_F$ oscillations corresponding to $2k_{Fl}$ are actually observed.
Figure \ref{FSFO} show that the $2k_{Fl}$ oscillations corresponding to 
the large Fermi surface are always dominating in the paramagnetic 
phase for various coupling constants and various densities. 

The Friedel oscillations obtained by the DMRG method clearly indicate
that the Fermi surface of the Kondo lattice model is large.  
At the early stage, the conclusions of the bosonization studies were 
controversial.  In the area of the paramagnetic metallic phase
Fujimoto and Kawakami obtained the Tomonaga-Luttinger liquid with a 
large Fermi surface, while White and Affleck predicted a
Luther-Emery liquid with a spin gap\cite{fujimoto,whiteaff}. Later,
it was argued that an additional direct Heisenberg coupling between
the $f$-spins is necessary to stabilize the Luther-Emery liquid
\cite{sikkema}.  Furthermore, the existence of a gapless excitation
with the momentum of $2k_{Fl}$ is shown rigorously by the 
Lieb-Schultz-Mattis construction \cite{yamanaka}. 

\begin{figure}
  \epsfxsize=85mm \epsffile{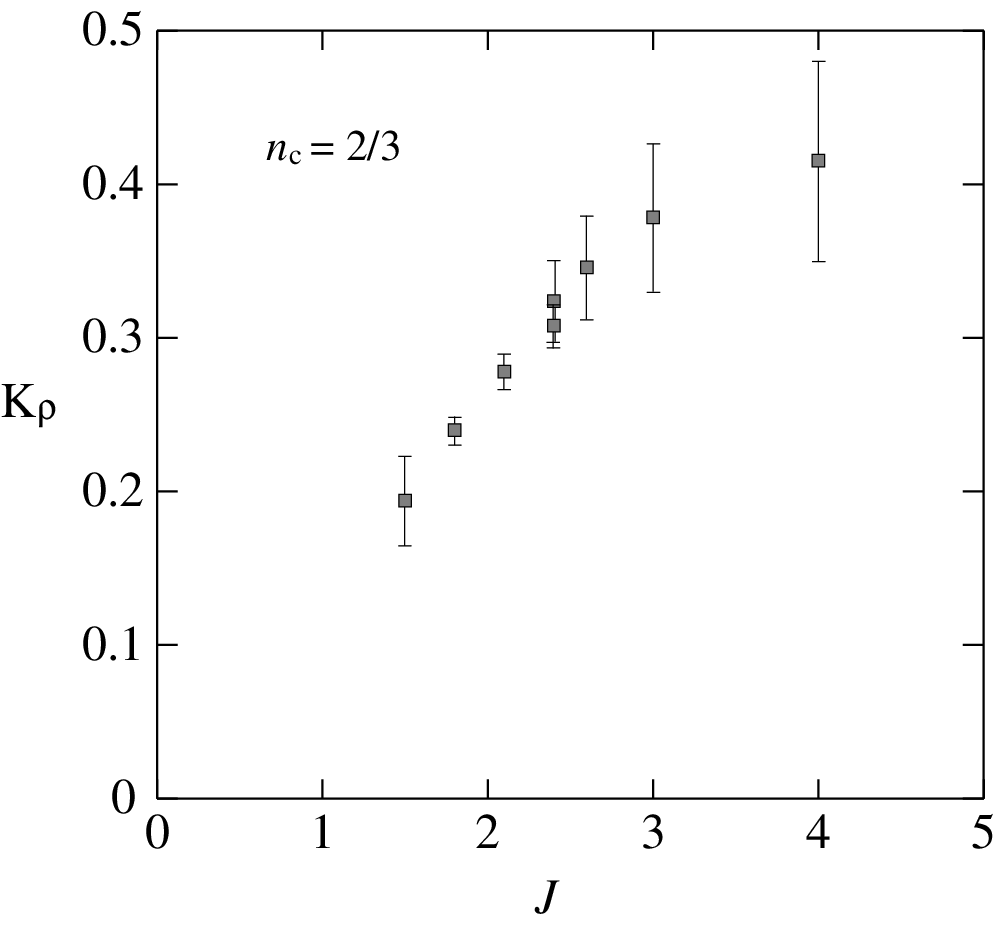}
\caption{
Correlation exponent $K_\rho$ estimated from the decay rate
of the charge density Friedel oscillations.
The errorbars are estimated from ambiguity of the
power law fitting. $n_c=2/3$. $J$ is in units of $t$.
}
\label{Kc}
\end{figure}

In order to obtain the correlation exponent $K_\rho$, we used the 
slope of the envelope function of the charge density oscillations,
assuming that dominant component of the oscillations is the $4k_F$
oscillations.  Figure \ref{Kc} shows $K_\rho$ thus determined for the 
exchange coupling constants from $J=4.0t$ to $1.5t$.  The density 
of conduction electons is fixed to $n_c=2/3$.  $K_\rho$ is always 
smaller than 1/2 and monotonically decreases with decreasing $J$.
The limiting value of $K_\rho=1/2$ in the strong coupling limit is 
easy to understand since the strong coupling limit of the Kondo 
lattice model is equivalent to the $U=\infty$ Hubbard model.

The correlation exponent shows a small discontinuity at the 
boundary between the ferromagnetic and paramagnetic phases,
$J_c = 2.4 t$ for $n_c=2/3$.  Below $J_c$, the $K_\rho$ decreases
faster and becomes lower than 1/3, which means that the long 
range behavior of the density-density correlation is governed by
the $4k_F$ oscillations rather than the $2k_F$ oscillations.
With further decreasing $J$, the $K_\rho$ seems to cross
the value $3-2\sqrt{2} \sim 0.17$.  Since the exponent of the 
power law anomaly of the momentum distribution function is 
given by $(K_\rho+ 1/K_\rho -2)/4$, the power law anomaly is 
removed below this point and a clear Fermi surface can not be 
seen anymore.  

Through the study of the Friedel oscillations by the DMRG method,
it has become clear that the paramagnetic metallic phase of the
one dimensional Kondo lattice model is a Tomonaga-Luttinger liquid
with a large Fermi surface.  This Tomonaga-Luttinger liquid is
unique in the sense that the $K_\rho$ is smaller than 1/2.
The small $K_\rho$ may be attributed to the long range nature of 
effective interactions with strong retardation
\cite{Schulz,Kolomeisky}. Recently, observation of the Friedel 
oscillations by the DMRG method is shown to be useful also for 
discussions of critical behaviors of the Hubbard model\cite{FoscHM}. 

\section{KONDO SPIN LIQUID PHASE AT HALF-FILLING}

The half-filled KL model is always insulating in one-dimension.
This conclusion was obtained by the 
exact diagonalization study \cite{THUS}.  Based on a finite
size scaling it was shown that finite excitation gap remains 
for any finite $J$. 
The result has been confirmed by the DMRG method \cite{Yu_White} and later 
supported by the mapping to a non-linear sigma model \cite{Tsve} and
by the bosonization approach \cite{Fujimoto2}. 
In this section we discuss basic properties of this insulating state.

Concerning the insulating phase, the question we would like to 
address here is which characters distintinguish the Kondo insulators
from usual semiconductors. The most significant difference is that
there are no gaps at high temperatures and they are induced as 
the temperature is lowered. Furthermore the excitation gaps induced 
by the temperature are different depending on channels. 
The difference of the excitation gaps and more generally the 
difference in the temperature dependence of various excitation 
spectra are naturally reflected in temperature dependence of 
various thermodynamic quantities.  
  
Clearly the lowest excitation gap which is the spin gap for the 
Kondo insulator defines the smallest energy scale of the system. 
In ordinary band insulators the band gap defines the smallest 
energy scale which controlls not only the spin excitations but 
also the charge excitations. It is also interesting to compare 
the smallest energy scale of the Kondo lattice problem with that
of the single impurity Kondo effect, namely the Kondo 
temperature $T_K$.  It is well known that the low temperature 
properties of the impurity model is governed by the single 
energy scale of $T_K$.

In the present section, first we will discuss the spin gap, the 
charge gap and the quasiparticle gap by the ground state DMRG. 
Then we will discuss temperature dependence of the spin 
susceptibility, the charge susceptibility and the specific heat 
by the finite-$T$ DMRG.  Temperature dependences of the single
particle excitation spectrum, the dynamic spin-spin correlation
function and the charge-charge correlation function are also 
discussed by the finite-$T$ DMRG,  For the analytic continuation
which is necessary to discuss the dynamic quantities, it is shown 
that the maximum entropy method is very useful\cite{MEM1,MEM2,MEM3}.

\subsection{Spin, charge and quasiparticl gaps}

To understand the physics of the insulating state of the 
half-filled Kondo lattice model, it is instructive to consider
the limit of strong exchange coupling $J$. 
In this limit every $f$ spin together with a conduction electron 
form a local singlet on every site.
To create spin excitations the minimum energy cost is $J$, which is the 
energy difference between the local spin singlet state and the local spin 
triplet state.  On the other hand, creation of charge excitations 
requires the minimum energy of $3J/2$ which corresponds to the 
energy cost for breaking two local singlets
by transferring a conduction electron to a neighboring site.

The excitation gaps monotonically decrease with decreasing 
exchange constant, but they do not vanish at any finite value of $ J $. 
Particularly, the weak coupling limit $ J \ll t $ is interesting.
In this regime the KL model is equivalent to the periodic Anderson model 
with strong Coulomb repulsion in the $f$ orbitals.
The salient feature of the strong Coulomb interaction
in the periodic Anderson model appears in the diverging ratio 
between the charge and spin gaps.
The limit of $J=0$ is singular where the conduction electrons 
and the $f$ spins are decoupled, and 
both the spin and charge gaps vanish.

For the discussion of the gaps, we take into account also the 
Coulomb interaction between the conduction electrons.
Since the spin and charge gaps are tiny in the weak coupling 
regime, it is no more justified to neglect the Coulomb
interaction between the conduction electrons. 
This Coulomb interaction suppresses the double occupation 
of conduction electrons, which eventually leads to the
formation of local magnetic moments of the conduction electrons.
Therefore the effect of the Coulomb interaction on the spin and 
charge gaps of the KL model sheds lights on the nature of the gap
formation in Kondo insulators.

The model we consider in this subsection is the following 
one-dimensional KL model with Coulomb interaction 
between the conduction electrons $U_c$:
\begin{eqnarray}
H & = & -t\sum_{i \sigma}
( c_{i \sigma}^\dagger c_{i+1 \sigma} + \mbox{H.c.})  
+J \sum_{i \mu} S^\mu_i \sigma^\mu_i \nonumber\\
 &  & + \ U_c\sum_i (c_{i \uparrow}^\dagger 
c_{i \uparrow}-\frac{1}{2}) (c_{i \downarrow}^\dagger 
c_{i \downarrow} -\frac{1}{2}) .
\end{eqnarray}
The Coulomb interaction is represented in the last term.
In this section  we consider the case of half-filling where the total 
number of conduction electrons is equal to the number of lattice 
sites $L$: $N_c\equiv \sum_{i \sigma} c^{\dagger}_{i \sigma}
c_{i \sigma}=L$.
This Hamiltonian is reduced to the Hubbard model
in the limit of $J \rightarrow 0$, and to 
the usual KL model for $U_c=0$.

In the impurity Kondo model all low temperature properties 
are scaled by the single energy scale $T_K \sim 
D\exp{(-\frac{1}{\rho J})}$,
where $\rho$ is the density of states 
of the conduction band at the Fermi level and is given by
$\frac{1}{2\pi t}$ in one dimension.
In contrast to the single impurity Kondo model, the KL model
has many $f$ spins which are coupled through the conduction 
electrons.
A basic question of the lattice problem is how the intersite
correlations appear in the energy scale.
The simplest extension of the form of $T_K$ may be an inclusion 
of an enhancement factor in the exponent:
The spin gap is expected to behave as 
\begin{equation}
\Delta_s \propto \exp{(-\frac{1}{\alpha \rho J})},
\label{s_gap}
\end{equation}
where $\alpha$ is the enhancement factor. 
The Gutzwiller approximation predicts an enhancement factor 
$\alpha=2$\cite{Rice}. 
Tsunetsugu $et\ al.$\  have\ estimated 
that the enhancement factor in one dimension is in the range of 
$1\le \alpha \le 5/4$ by using a finite size scaling for the  
results obtained by the exact diagonalization\cite{THUS}. 
In the following we present the results on the enhancement factor
obtainedy by the DMRG. 

\begin{figure}
  \epsfxsize=85mm \epsffile{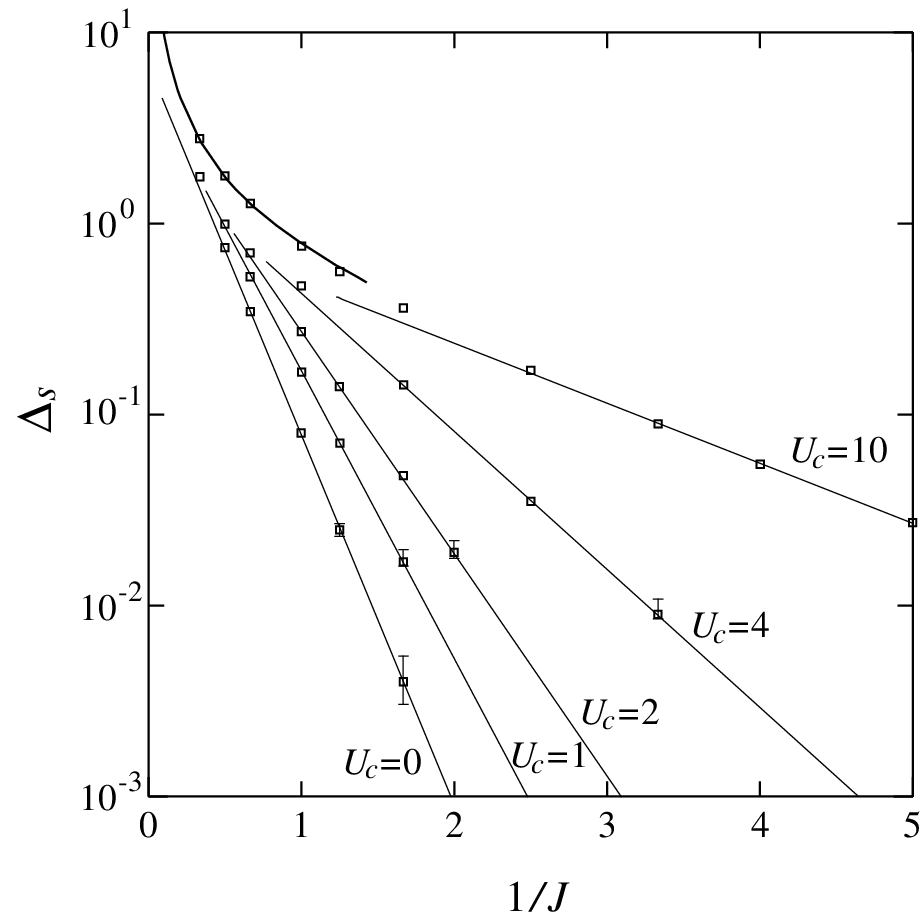}
\caption{
Spin gap of the half-filled one-dimensional Kondo lattice model
with Coulomb interaction. The thick curve represents the 
result of the perturbation theory in terms of $t/J$ for $U_c=10t$.
Typical truncation error in the DMRG calculation is 
$10^{-6}$ for $J=1$.
Errorbars are estimated from $L^{-1}$ and $L^{-2}$ scalings. 
Gap energies, exchange constant $J$, and 
Coulomb interaction $U_c$ are in units of $t$.
}
\label{spin_gap}
\end{figure}

The spin gap is obtained from the 
difference of the ground-state energies in the subspaces of 
total $S^z$ being zero and one, Eq.(\ref{spin_gap_def});
the SU(2) symmetry in the spin space guarantees the energy 
difference is the same as the spin gap in the subspace of 
zero total $S^z$. 
The spin gap of the bulk system is estimated from the following 
scaling function:
\begin{equation}
\Delta_s(L)=\Delta_s(\infty)+\beta L^{-2}+O(L^{-4}).
\label{finite}
\end{equation}
The obtained spin gaps are plotted in Fig. \ref{spin_gap}
in logarithmic scale as a function of $1/J$.
The results are obtained by the extrapolation to the bulk limit 
using data of $L=6,8,12,18,24,40$. The DMRG calculations were 
done by using the finite system algorithm with 
open boundary conditions keeping up to 300 states for each block.
The enhancement factor is obtained from the slope in 
the figure, and determined to be $\alpha  =1.4(1)$ 
for $U_c=0$.
There are some uncertainties in the extrapolation 
to the bulk limit for tiny gaps. However within the present 
accuracy we do not observe any indication of the logarithmic 
correction to the exponent which was predicted by the 
semiclassical approach\cite{Tsve}.

\begin{figure}
  \epsfxsize=85mm \epsffile{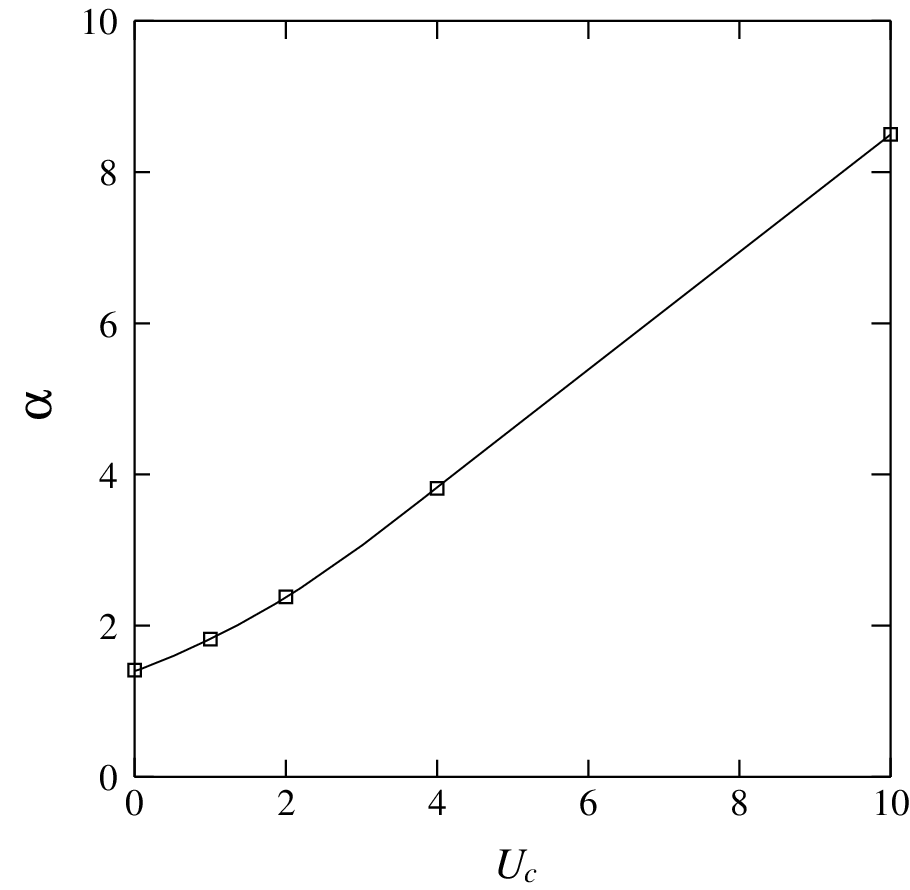}
\caption{
$U_c$-dependence of the exponent of the spin gap.
Coulomb interaction $U_c$ is in units of $t$.
}
\label{alpha}
\end{figure}

Now we consider the effect of the Coulomb interaction.
In the weak coupling region it is natural to extend the 
form of Eq.\ (\ref{s_gap}) to finite $U_c$ allowing 
$U_c$-dependence of the exponent.
Indeed the numerical data are nicely fitted by this form
as shown in Fig.\ \ref{spin_gap}.
The obtained $U_c$-dependence is shown in Fig.\ \ref{alpha}, 
which indicates that $\alpha(U_c)$ increases with increasing 
$U_c$ and the asymptotic behavior is linear in $U_c$.
The KL model with the Coulomb interaction is mapped to 
a Heisenberg chain coupled with the localized $f$ spins in 
the limit of $U_c/t \rightarrow \infty$.
The linear $U_c$-dependence of the exponent $\alpha = 0.78U_c/t 
+ 0.7$ in Fig.\ \ref{alpha}
means that the spin gap of the effective spin model behaves as 
$\Delta_s \sim \exp{(-2J_{\mbox{\scriptsize eff}}/J)}$
with $J_{\mbox{\scriptsize eff}}=4t^2/U_c$ beeing
the effective coupling of the Heisenberg chain.
In order to check this form we have analyzed
the numerical data for the spin system obtained 
by Igarashi $et\ al.$ \cite{Igarashi}, 
and found a good coincidence.
Thus we conclude that the enhancement factor $\alpha$ increases 
monotonically with increasing $U_c$.
This is natural since the origin of the spin gap is the singlet 
binding between the localized spins and the conduction electons
for which the Coulomb interaction helps by suppressing the 
double occupancy.

In contrast to the single impurity Kondo model, the KL model has 
the second energy scale that characterizes the charge excitations at
low temperatures. 
The charge excitations keep spin quantum numbers,
and the charge gap is defined by the difference of the
lowest energy in the subspace of $N_c=L$ and $N_c=L+2$: 
$E_g(L,N_c=L+2,S=0)-E_g(L,N_c=L,S=0)$.
Owing to the hidden SU(2) symmetry in the charge space, 
the energy difference is the same as the charge excitation gap
in the subspace of the fixed number of electrons $N_c=L$ 
\cite{Nishino}.

\begin{figure}
  \epsfxsize=85mm \epsffile{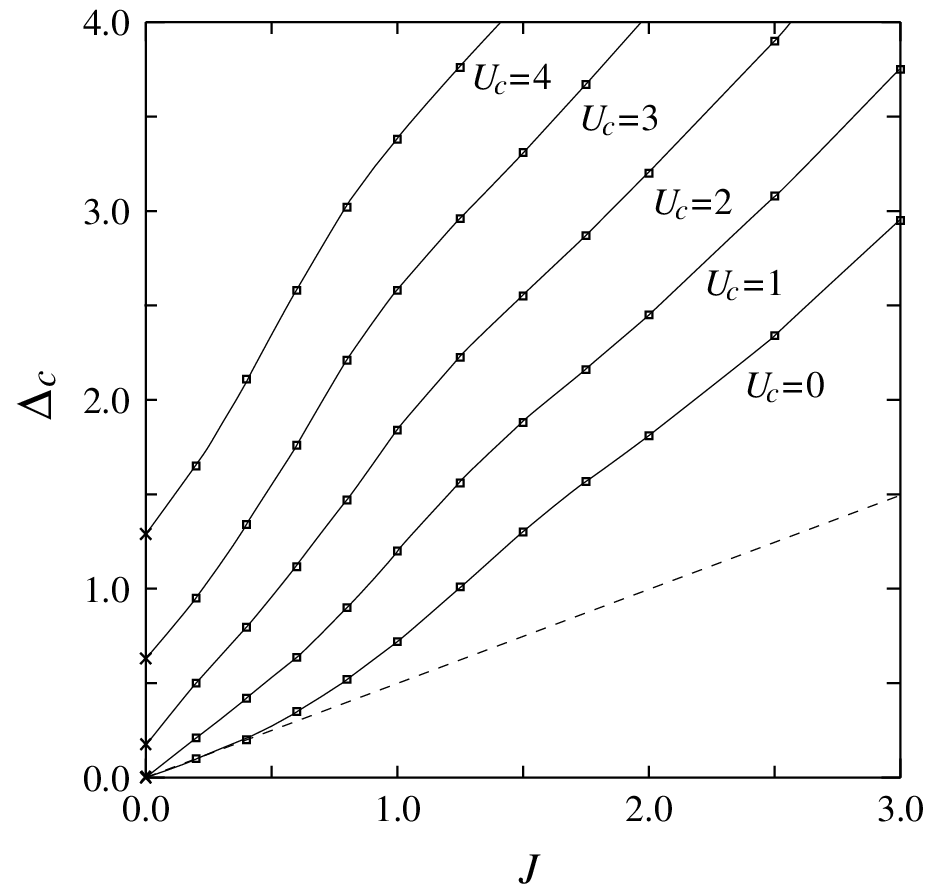}
\caption{
Charge gap of the half-filled one-dimensional Kondo lattice model
with Coulomb interaction. Results on the vertical axis are 
obtained from the exact solution by Lieb-Wu.
Typical truncation errors in the DMRG calculation are 
$10^{-6}$ for $J=1$ and $10^{-4}$ for $J=0.2$, 
which are dominant source of numerical errors since
the finite size scaling, Eq.\ (\ref{finite}), is well obeyed.
Gap energies, exchange constant $J$, 
and Coulomb interaction $U_c$ are in units of $t$.
}
\label{charge_gap}
\end{figure}

Figure \ref{charge_gap} shows the charge gap obtained by the 
extrapolation to the infinite system.
The results for $J=0$ are known as the Hubbared gap
of the one-dimensional Hubared model which is exactly solved 
by the Bethe Ansatz \cite{Lieb}. 
The asymptotic forms of the charge gap are given by 
$\Delta_c\propto \sqrt{U_ct}\exp
{(-\frac{1}{\rho U_c})}$ for small $U_c$,
and by $\Delta_c\propto U_c-4t$ for large $U_c$.
The results obtained for finite $J$
are consistent with the exact ones which are 
denoted by the crosses on the vertical axis.

For $U_c=0$ the charge gap is linear in $J$ in the small $J/t$ limit.
As is shown by the exact diagonalization study
the charge gap is much bigger than the 
spin gap in the weak coupling regime \cite{Nishino}.
It implies that the correlation length for the spin degrees of 
freedom is much longer than the charge correlation length. 
Therefore for the discussion of the charge gap it is 
justified to assume that the spin-spin correlation 
length is infinitely long.
Under the assumption of the infinite spin correlation length,
the charge gap is calculated as 
\begin{equation}
 \Delta_c = \frac{J}{2}.
\end{equation}
From Fig.\ \ref{charge_gap} we also find that 
the charge gap increases with increasing Coulomb interaction $U_c$.

The charge excitations are created by adding two additional electrons
keeping spin quantum numbers fixed. When we put single electron 
in the ground state, then a quasiparticle excitation is made.
Here we consider the relation between the charge gap $\Delta_c$
and the quasiparticle gap $\Delta_{qp}$
which is defined by $E_g(L,N_c=L\pm 1,S=\pm 1/2)-E_g(L,N_c=L,S=0)$.
In the strong coupling limit, $J/t \rightarrow \infty$, 
it is evident that the charge gap is twice
the quasiparticle gap 
owing to the SU(2) symmetry in the charge space. 
In the second order perturbation in $t/J$, one can show 
that the interaction between the two additional electrons is 
repulsive, leading to only a phase shift. 
Therefore the charge gap in the bulk limit is twice 
the quasiparticle gap $\Delta_{qp}$:
\begin{equation}
\Delta_c=2\Delta_{qp}.
\end{equation}
A similar argument is also valid for the periodic 
Anderson model \cite{Nishino}.
Validity of this relation is checked by the DMRG calculation 
in the entire region of the exchange constant $J$. 
Concerning the spin gap, the lowest spin excitation 
may be considered as a
bound state of a quasielectron and a quasihole.

\subsection{Susceptibilities at finite temperatures}
The spin and charge gaps determined at zero temperature are
very different in the weak coupling regime. 
The spin gap is exponentially small, 
while the charge gap is proportional to $J$.
The large charge gap originates from the staggering
internal magnetic fields induced by the long correlation 
length of the $f$ spins. 
At finite temperatures, however, the spin correlations are subject
to the thermal fluctuations.
When temperature becomes comparable to the spin gap, the spin 
correlation length gets smaller and the whole electronic states
including the charge excitation spectrum are reconstructed. 
In this section we study such an interplay between the 
spin and charge excitations at finite temperatures by looking 
at the thermodynamic quantities.  In what follows we consider
the original KL model neglecting the Coulomb interaction 
between the conduction electrons.

In order to calculate thermodynamic quantities we use the 
finite-$T$ DMRG discussed in Sec.\ {\bf II} \cite{shibata,wang}.
In this method the free energy is obtained from 
the maximum eigenvalue of the quantum transfer matrix. 
The spin and charge susceptibilities are obtained from the 
derivatives of the free energy with respect to external magnetic 
field or chemical potential.
The calculations are performed by the infinite system 
algorithm keeping 40 states per block.
The truncation errors in the calculation are
typically $10^{-3}$ and at the lowest temperature $10^{-2}$
for the Trotter number $M=50$.

\begin{figure}
  \epsfxsize=85mm \epsffile{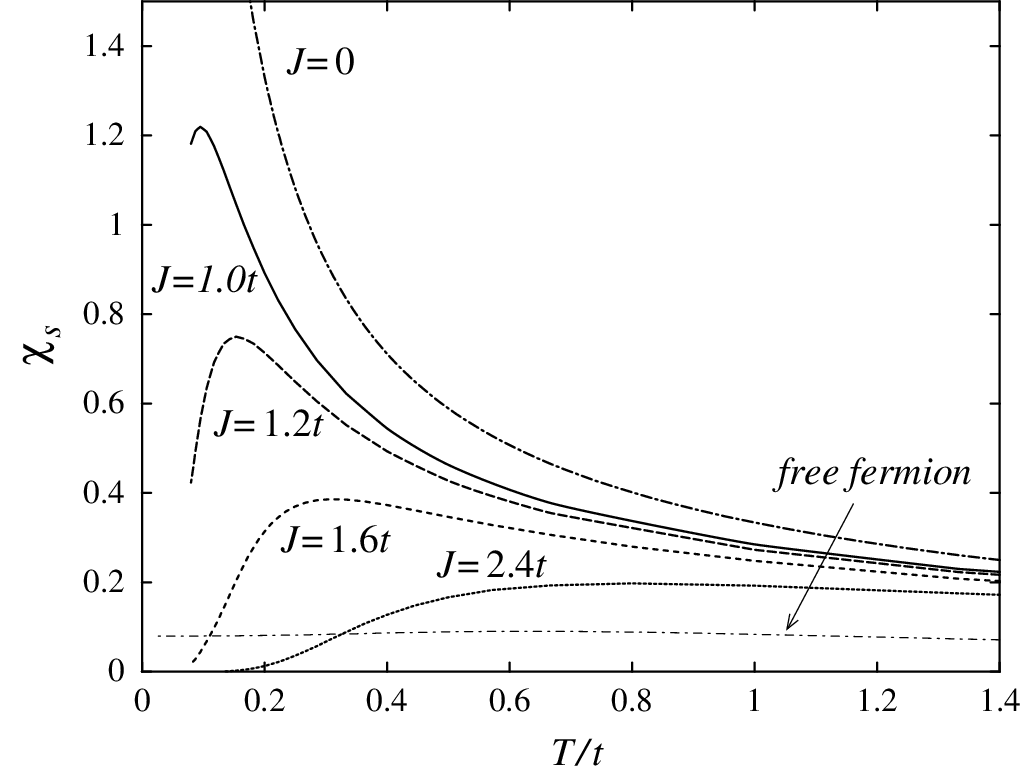}
\caption{
Spin susceptibility of the half-filled 
one-dimensional Kondo lattice model. The truncation 
errors in the finite-$T$ DMRG calculation are
typically $10^{-3}$ and at the lowest temperature $10^{-2}$.
}
\label{spin_sus}
\end{figure}

We first consider the temperature dependence of the uniform
spin susceptibility. 
The spin susceptibility is obtained from the change 
of the free energy by a small magnetic field $h$: 
$\delta F=\chi_s h^2/2$
The results for $J/t=0, 1.0, 1.2, 1.6$ and $2.4$ 
are shown in Fig. \ref{spin_sus}.

When $J/t=0$, the localized spins and 
conduction electrons are uncorrelated.
The susceptibility is given by the sum of the Curie term 
due to the free $f$ spins and the Pauli term of the free 
conduction electrons. The contribution of the Pauli 
susceptibility of the conduction electrons shown by the dashed line 
in Fig. \ref{spin_sus}
is relatively small, and the total susceptibility for $J/t=0$ 
is dominated by the Curie term.

For finite $J$, the low temperature part of $\chi_s$ 
sharply drops with decreasing the temperature.
This drastic change is due to the appearance of a low energy scale
for the spin sector.
The spin gap of $0.08t$ for $J/t=1.0$ is consistent with
the characteristic temperature at which $\chi_s$
starts to decrease deviating from the Curie low.

In order to determine the energy scale at low temperatures 
we estimate the activation energy by fitting the 
susceptibility with an exponential form. 
The estimated activation energy for the spin susceptibility
is summarized in Table \ref{table-2} for $J/t=1.6, 2.4$ and $ 3.0$. 
Compared with the quasiparticle gap and the spin gap, both of which are 
responsible for the magnetic excitations, we conclude that the lower
one of them determines 
the low-temperature energy scale of the spin susceptibility.
This is consistent with the general form of
susceptibility which is written as 
\begin{eqnarray} 
    \chi_s &=& Z^{-1} N^{-1} \beta \sum_{m} {\rm e}^{-\beta E_m}
    {\langle m | S_z^{\rm total} | m \rangle} ^2\  ,\label{sus_eq}\\
       Z &=& \sum_m {\rm e}^{-\beta E_m}.\label{Z_eq}
\end{eqnarray}
The point is that Eqs.(\ref{sus_eq}) and (\ref{Z_eq})
apply for both the canonical 
and the grand canonical ensembles by properly defining 
the states $| m \rangle$. In the thermodynamic 
limit the susceptibilities by the two ensembles should 
give the same answer. From this consideration it is 
concluded that the smaller one of the spin gap and 
the quasiparticle gap determines the low temperature
energy scale.
For the case of small exchange coupling $(J/t \ll 1)$
the spin gap is smaller than the quasi-particle gap and 
thus the low temperature energy scale of $\chi_s$ is 
determined by the spin gap.

\begin{figure}
  \epsfxsize=85mm \epsffile{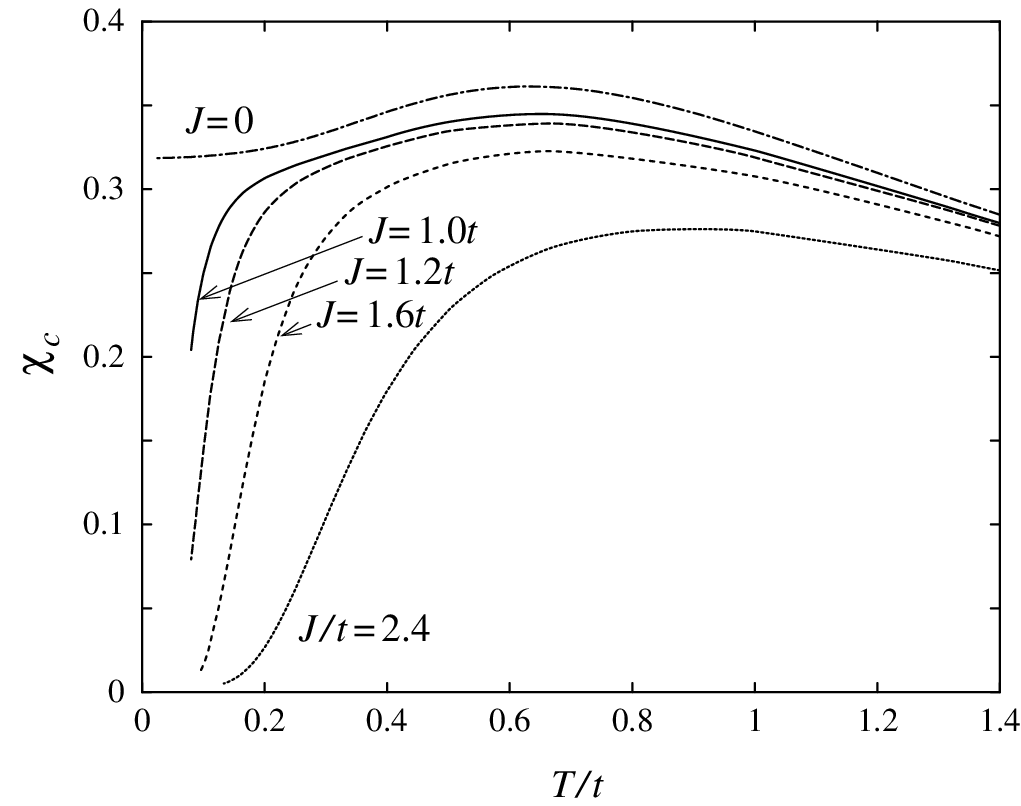}
\caption{
Charge susceptibility of the half-filled 
one-dimensional Kondo lattice model. 
}
\label{charge_sus}
\end{figure}

In order to see the effect of thermal fluctuations 
of $f$ spins to the charge excitations,
we next calculate the charge susceptibility $\chi_c$. 
The $\chi_c$ is obtained from the change of the free 
energy due to a small shift of chemical potential $\mu$,
$\delta F=\chi_c \mu^2/2$. In the present calculation 
we use the fact that 
the chemical potential is zero at half-filling 
owing to the SO(4) symmetry of the model.\cite{Nishino} 
The results for $J/t=0, 1.0, 1.2, 1.6$ 
and $2.4$ are shown in Fig. \ref{charge_sus}. 

For $J/t=0$, $\chi_c$ does not show diverging 
behavior at low temperatures in contrast to $\chi_s$. 
In the limit of $T=0$, $\chi_c$ is equal to 
the density of state of conduction electrons
which is $1/\pi t$ including the two spin directions.
This is expected since the charge degrees of freedom is 
governed by the conduction electrons.
Since there is no interaction, 
$\chi_c/4$ is equal to the spin susceptibility 
of the free conduction electrons.
The slight increase in $\chi_c$ in the low-temperature
region is a characteristic feature of the
one-dimensional system where the density of states 
diverges at the band edges. 

A finite value of $J$ produces a sharp drop 
in $\chi_c$ at low temperatures. 
Similarly to $\chi_s$, this drop is due to the appearance 
of low energy scale $\Delta_c$ for the charge sector. 
The energy scale is determined
by the activation energy for the charge susceptibility.
By fitting $\chi_c$ with an exponential form,
the activation energy $\Delta_{\chi_c}$ is obtained 
as listed in Table \ref{table-2} for $J/t=1.6, 2.4$ and $ 3.0$.
From this table it is concluded that the quasiparticle gap 
determines the low-temperature energy scale of the charge
susceptibility.

\begin{table}
\caption{
Activation energy obtained from the 
spin and charge susceptibilities, $\Delta_{\chi_s}$ and 
$\Delta_{\chi_c}$, of the one-dimensional Kondo lattice model
at half-filling. The quasiparticle gap $\Delta_{qp}$ and 
the spin gap $\Delta_s$ are obtained by the zero-temperature
DMRG.
The charge gap is twice the quasiparticle gap;
$\Delta_c = 2\Delta_{qp}$ 
}
\begin{tabular}{lcccc} 
 & $\Delta_{\chi_s}/t$ & $\Delta_{\chi_c}/t$ & $\Delta_s/t$ 
& $\Delta_{qp}/t$ \\ \hline
$J/t=1.0$ &  & & $0.08 $ & $ 0.36$  \\
$J/t=1.2$ &  & & $0.16 $ & $ 0.47$  \\
$J/t=1.6$ & $0.45 \pm 0.1$& $0.6 \pm 0.1$ & $0.4 $ & $ 0.7$  \\
$J/t=2.4$ & $1.2 \pm 0.1$ & $1.0 \pm 0.1$ & $1.1 $ & $ 1.1$  \\
$J/t=3.0$ & $1.6 \pm 0.1$ & $1.4 \pm 0.1$ & $1.8 $ & $ 1.5$  \\ 
\end{tabular}
\label{table-2}
\end{table}

Although the qusiparticle gap determine the exponential 
temperature dependence at low temperatures, it is not the single 
energy scale for $\chi_c$.  It may be best understood 
by looking at $\chi_c$  for $J/t=1.0$.  A sharp decrease
of $\chi_c$ is seen at around $T\sim0.1t$ which is much
smaller than the quasiparticle gap $0.36t$ but rather 
close to the value of the spin gap $0.08t$. This fact 
suggests that the whole electronic states are reconstructed
when temperature is raised up to the spin gap. 
We will discuss this aspect in more detail in connection 
with the temperature dependence of various excitation 
spectra.  

\subsection{Specific heat}

\begin{figure}
  \epsfxsize=85mm \epsffile{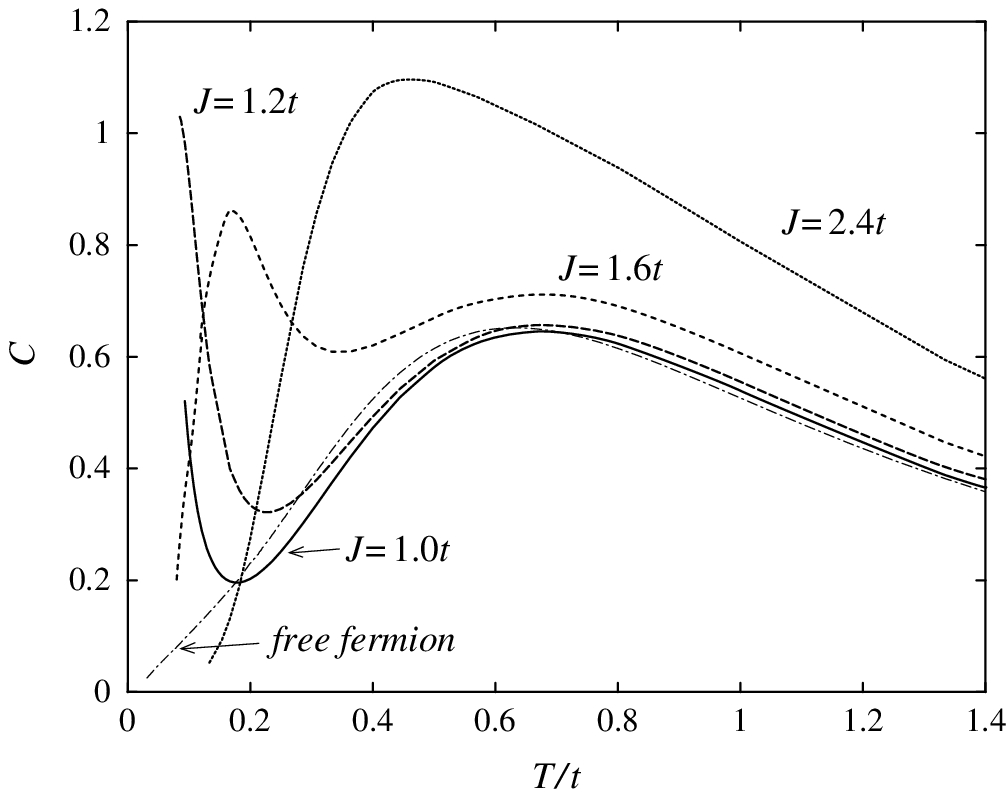}
\caption{
Specific heat of the half-filled 
one-dimensional Kondo lattice model. 
}
\label{specific_heat}
\end{figure}

In order to see how the entropy of the system is
released, we next calculate specific heat.
The specific heat is calculated from the second derivative 
of the free energy; $C=-T\partial^2 F/\partial T^2$.
The results for $J/t=0, 1.0, 1.2, 1.6$ and $2.4$ are shown in 
Fig. \ref{specific_heat}.

At $J/t=0$ the specific heat of this model is given by the 
sum of the delta function at $T=0$ that originates from the 
free localized 
spins and the specific heat of free conduction electrons.
For finite $J$ they are combined to form a two-peak structure. 
The peak at higher temperatures is almost independent of 
the exchange constant and similar to the specific heat of free 
conduction electrons. 
Thus the structure at higher temperatures may be understood as
the band structure effect of the one dimensional conduction electrons.
In contrast to the higher temperature structure,
the structure at lower temperatures strongly changes its form
with $J$.
The peak shifts towards higher temperatures and becomes 
broader with increasing $J$.

With further increasing the exchange coupling.
the spin gap becomes comparable to the hopping
matrix element $t$.
In this situation various energy scales are not distinguishable
and the specific heat possesses a single peak structure 
as shown for $J/t=2.4$.

\subsection{Dynamic properties}
Dynamic properties of the half-filled KL model show
clear features characteristic to strongly correlated insulators.
Unusual temperature dependence of the excitation spectra 
is one of the most important features of the interacting systems.
Actually the behaviors of the static susceptibilities 
discussed in the preceding subsection indicates that the 
excitation gaps develop at low temperatures.

In this section we calculate the dynamic spin and charge 
structure factors, $S(\omega)$ and $N(\omega)$, and the density 
of states, $\rho(\omega)$.  By looking at their temperature 
dependence we can study temperature evolution of the excitation 
gaps and the relations among dynamic quantities.
To obtain the dynamic quantities we first calculate the correlation 
functions in the imaginary time direction.
As we have discussed in section II,  the correlation functions 
in the imaginary time direction are directly calculated from
the left and right eigenvectors obtained by applying the 
finite-$T$ DMRG to the transfer matrix.

\begin{figure}
  \epsfxsize=85mm \epsffile{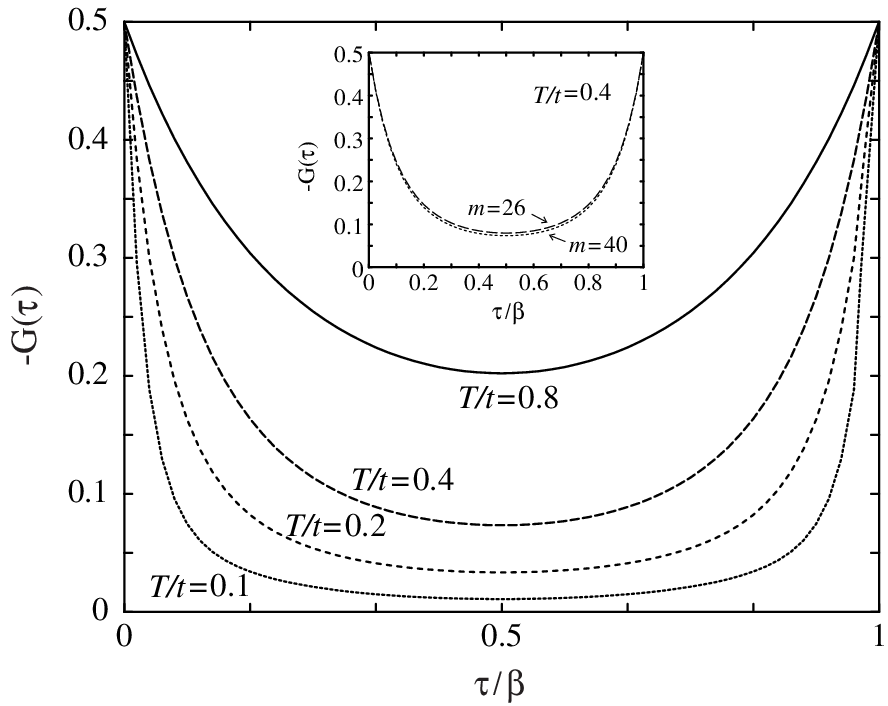}
\caption{
Imaginary time correlation functions, $G(\tau)$,
of the half-filled one-dimensional Kondo lattice model: 
$J/t=1.6$. The Trotter number $M=60$.
Inset shows the results for diffenent number of state 
kept $m$ in the DMRG calculation.
The truncation errors in the DMRG calculation are
$ 2 \times 10^{-3}$ for $m=26$ and 
$ 3 \times 10^{-4}$ for $m=40$.}
\label{imag_corr}
\end{figure}

Examples of the single particle Green's function
as a function of the imaginary time calculated by the 
finite-$T$ DMRG are shown in Fig. \ref{imag_corr}.
To obtain the spectral functions we first Fourier transform 
the imaginary time correlation functions and then need to perform 
analytic continuation from the imaginary frequency axis 
to the real frequency axis. One straightforward method for 
the analytic continuation is to use the Pad\'e approximations. 
Since the DMRG calculation yields no statistical errors
the Pad\'e approximations show good convergence in many cases.
However, it is still difficult to obtain the spectral functions
by the Pad\'e approximations in a stable manner when a spectrum
has a nearly singular form.  The reason is that the Pad\'e approximations 
use rational functions of Matsubara frequencies $i\omega_n$.

\begin{figure}
  \epsfxsize=85mm \epsffile{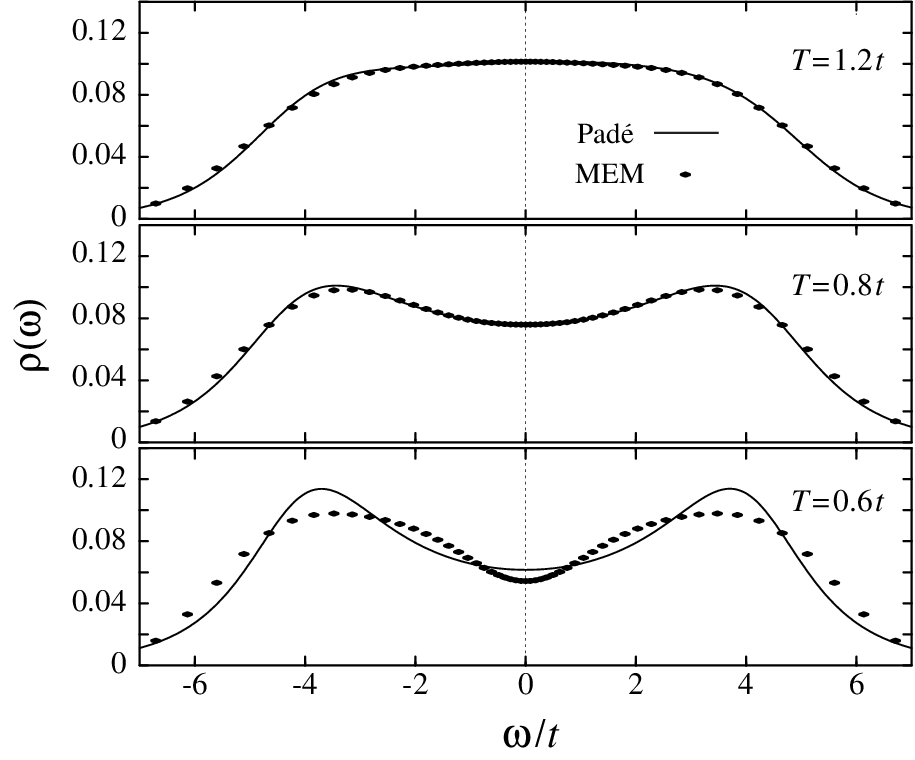}
\caption{
Quasiparticle density of states, $\rho(\omega)$,
obtained from the $G(\tau)$ by the Pad\'e approximations
and the maximum entropy method (MEM). The Trotter number $M=60$.}
\label{Pade_MEM}
\end{figure}

Even when a spectrum has a nearly singular form, 
the maximum entropy method still works well. 
An advantage of this method is that we can 
explicitly use the symmetries and the positiveness of the spectral function.
The results obtained by the two method are compared in
Fig. \ref{Pade_MEM}. At high 
temperatures two results coincide with each other, but with decreasing the 
temperature the convergence of the Pad\'e approximations becomes worse 
due to the growing  
singularity in the spectral function in the low frequency region. 
The results by the maximum entropy method are stable even at low 
temperatures. 
Thus in the following we employ the maximum entropy method to 
study temperature 
evolution of the dynamic correlation functions\cite{Mutou}.

\begin{figure}
  \epsfxsize=85mm \epsffile{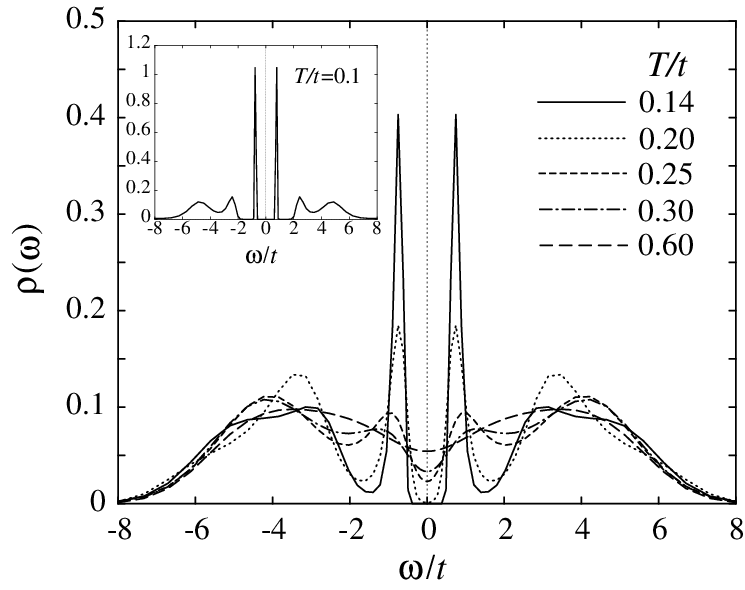}
\caption{
Quasiparticle density of states, $\rho(\omega)$,
of the half-filled one-dimensional Kondo lattice model: 
$J/t=1.6$.
The Trotter number $M=60$, and the number of states kept $m=40$. 
}
\label{dos}
\end{figure}

The quasiparticle density of states obtained for $J/t=1.6$ at temperatures 
$T/t=0.1,\,0.14,\,0.2,\,0.25,\,0.3,\,0.6$ is shown in Fig. \ref{dos}.
Existence of the quasiparticle gap is seen as a clear
dip structure around $\omega=0$.
At low temperatures sharp peaks appear at $\omega=\pm\Delta_{qp}$
separated from the higher frequency part of the
spectral weight. 
Sharpness of the peaks suggests
the formation of the heavy quasiparticle bands at the gap edges.
The high frequency part of the spectral weight
extends to the region far from
the edge of the free conduction band $\omega=2t$,
which shows significance of the multiple excitations accompanied
with the quasiparticle excitations.

As the temperature is increased, the peak at the threshold 
gets broadened and at temperatures $T\sim\Delta_s$ the peak
and the accompanied dip between the low and high frequency parts 
completely disappear. 
This result shows that although the sharp peaks of the 
quasiparticle density of states are located at the frequency 
$\omega=\pm\Delta_{qp}$, it disappears around the temperature 
$T\sim\Delta_s$ which is much lower than $T\sim\Delta_{qp}$.

\begin{figure}
  \epsfxsize=85mm \epsffile{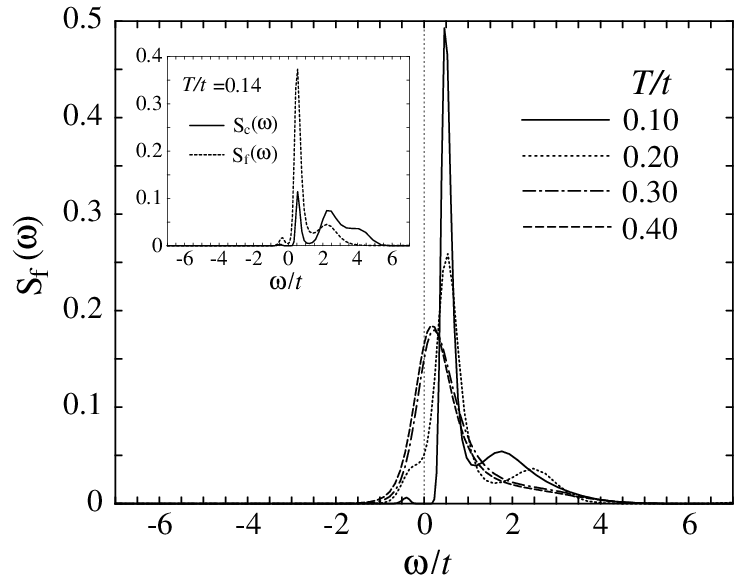}
\caption{
Dynamic spin structure factor of the $f$ spins, $S_f(\omega)$,
of the half-filled one-dimensional Kondo lattice model: $J/t=1.6$.
}
\label{S_f}
\end{figure}

In order to see what happens at the temperatures $T\sim\Delta_s$,
we next consider the $f$ spin dynamic structure factor $S_f(\omega)$. 
The calculated $S_f(\omega)$ are presented in Fig. \ref{S_f}.
At the lowest temperature the spin gap is clearly seen with a sharp peak
at the gap edge. This characteristic peak has the most of 
the spectral weight, which shows concentrated $f$ spin excitations
at the energy scale of $\Delta_s$. There is a broad peak 
in the higher frequency side. As is shown later, a similar structure 
and temperature dependence appear in the dynamic spin structure 
factor of the conduction electrons $S_c(\omega)$. 
Thus we conclude that through the exchange coupling
excitations of the $f$ spins are mixed with those of conduction spins,
which yields this broad peak in higher frequency part of $S_f(\omega)$.

With increasing the temperature, the peak structure at $\omega=\Delta_s$
becomes broad and the spectral intensity increases
around the zero frequency $|\omega|<\Delta_s$.
At the temperatures $T\sim\Delta_s$,
the peak position of the spectrum shifts to the zero frequency,
and the peak height becomes almost temperature independent.
The spectral intensity at the zero frequency 
is directly related to the NMR relaxation rate $1/T_1$.
Hence the present results show that
$1/T_1$ is nearly temperature independent at high temperatures and
drastically decreases with decreasing temperature
below the characteristic temperature of the order $\Delta_s$.

\begin{figure}
  \epsfxsize=85mm \epsffile{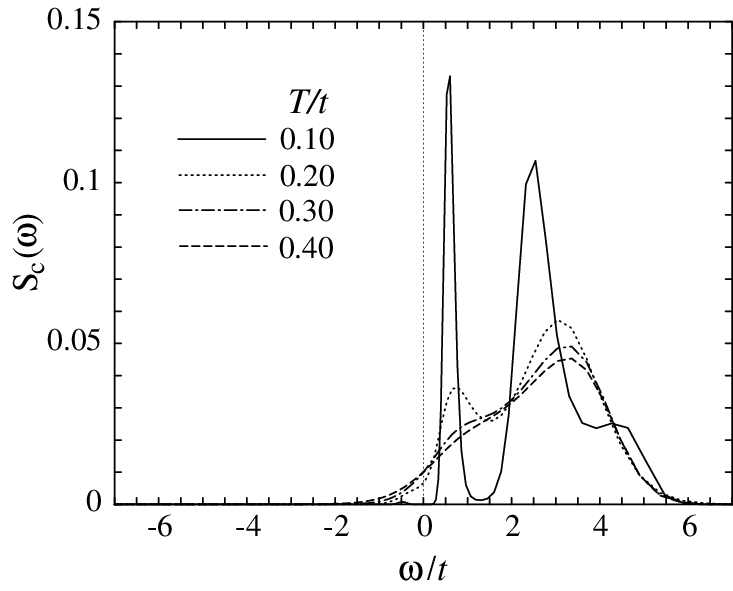}
\caption{
Dynamic spin structure factor of the conduction electrons, $S_c(\omega)$,
of the half-filled one-dimensional Kondo lattice model: 
$J/t=1.6$.
}
\label{S_c}
\end{figure}

The dynamic spin structure factor for the 
conduction electrons, $S_c(\omega)$, is shown in Fig. \ref{S_c}.
At low temperatures, $S_c(\omega)$ has two peaks.
The peak in the low frequency side is located at the energy
of $\Delta_s$ similarly to $S_f(\omega)$.
This peak corresponds to the spin excitations of
the singlet bound states of conduction electrons with $f$ spins
to triplet states.
The high frequency peak is located slightly above the charge gap,
that corresponds to the spin excitations due to quasiparticles.
With increasing the temperature both peaks lose
their intensity, and above the temperature $T\sim\Delta_s$ 
the low frequency peak structure disappears.
The spectrum at the high frequency side becomes similar to
that of the dynamic charge structure factor $N(\omega)$.
This means that high frequency excitations are dominated by
the quasiparticle excitations of almost free conduction 
electrons and thus the relation $S_c (\omega)=N_c (\omega)/4$
is satisfied approximately.

\begin{figure}
  \epsfxsize=85mm \epsffile{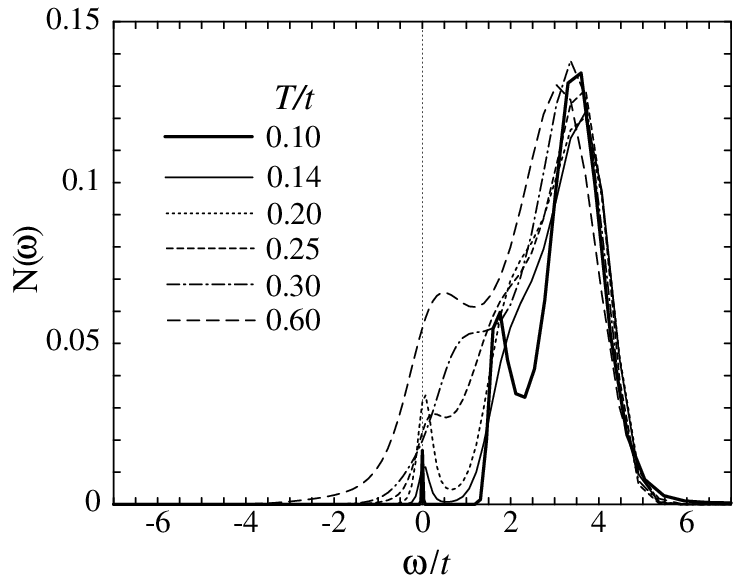}
\caption{
Dynamic charge structure factor of conduction electrons, $N_c(\omega)$,
of the half-filled one-dimensional Kondo lattice model: 
$J/t=1.6$.
}
\label{N_c}
\end{figure}

The dynamic charge structure factor $N(\omega)$ is shown in 
Fig. \ref{N_c}.
At the lowest temperature two clear peaks appear,
a smaller peak at $\omega\sim 0$ and 
a bigger one at $\Delta_c$.
These two peaks originate from the sharp peak structure
in $\rho(\omega)$ at $\omega=\pm\Delta_{qp}$.
The excitations of thermally populated quasiparticles within 
the sharp peak in $\rho(\omega)$ contribute to the peak at $\omega=0$,
while the excitations between the peaks in $\rho(\omega)$
make the peak in $N(\omega)$ at $\omega\sim\Delta_c$.
With increasing temperature, increased number of the thermally 
populated quasiparticles enhances the peak at $\omega=0$,
but at the temperature $T\sim\Delta_s$
the peak structure is completely smeared out, which reflects the 
disappearance of the peak in $\rho(\omega)$. 
The gap structure of $N(\omega)$ 
and the energy scale of $\Delta_c$ become unclear
at temperatures much smaller than $\Delta_c$.

Dynamic quatities studied by the finite-$T$ DMRG have revealed 
the many-body nature of the gap formation in the Kondo insulators.
The difference among excitation gaps depending on channels is 
a characteristic feature of the Kondo insulators compared with 
the ordinary band insulators.  The temperature induced gap 
formation for the single particle density of states is another
clear evidence of the many-body feature.  
Even at a fixed temperature a renormalized 
band picture fails to capture the essential physics of the strongly
correlated insulators.  A typical example is that the two-body 
excitation spectrum $N(\omega)$ is very different from a 
convolution of the one-body excitation spectrum $\rho(\omega)$.
In the Kondo insulators there are several low energy scales, 
corresponding to the spin gap, the quasiparticle gap and the charge gap.
Among them the lowest one, the spin gap, plays a special role.
At higher temperatures than the spin gap, the excitation spectra 
in the charge sector are also modified strongly.  It means that the
whole electronic states are reconstructed above the tempeature
corresponding to the lowest energy scale. 

\section{SUMMARY AND DISCUSSIONS}

In this review we have discussed Tomonaga-Luttinger liquid 
properties of the one-dimensional Kondo lattice model away 
from half-filling. In particular, the large Fermi surface 
is concluded in the ground state by investigating the 
spin and charge Friedel oscillations. 
At half-filling of the one-dimensional Kondo lattice model 
the ground state is always an incompressible spin liquid 
phase. Studies on the dynamic correlation functions have revealed
many-body nature of this insulating phase in several ways.

These developments have been achieved by applying the density matrix 
renormalization group method either to the Hamiltonian itself
or to the quantum transfer matrix.  In the problem of Kondo lattice 
model there appear small energy scales at low temperatures.  
It implies that the correlation lengths for various quantities 
are relatively long and therefore we need sufficiently long 
systems to observe intrinsic properties.  On the other hand, 
there are 8 states per site in the Kondo lattice problem.  
Of course for a quantum spin-1/2 chain there are only two states 
per site.  Exact diagonalization studies can 
do a good job for the latter but only poor one for the former.
In this situation the DMRG shows its full advantage for the Kondo 
lattice model.  

We would like to stress that now we can calculate dynamic 
quantities at finite temperatures by applying the finite-$T$
DMRG to the quantum transfer matrix.  This method is 
free from statistical errors and the truncation errors 
are the only source of numerical errors.  Therefore 
much better accuracy is obtained for the imaginary time
data, from which corresponding spectral function may be
obtained reliably through the maximum entropy method.
Another advantage of the finite-$T$ DMRG compared with the 
quantum Monte Carlo simulations is that we do not have the 
negative sign problem for any quantum systems.  
 
Generally speaking, more elaborate calculations are required 
for the finite-$T$ DMRG compared with the zero-temperature 
DMRG.  Finite temperature properties, both static and dynamic, 
of the Tomonaga-Luttinger liquid phase
are of great interest and studies in this direction are now
being in progress.  In near future it will become possible to
address these questions.
Investigation of a completely Fermionic model for heavy Fermions,
for example the periodic Anderson model, is also left for future.
 
%
%
%
    
\acknowledgments
It is our great pleasure to acknowledge fruitful collaborations with
Tetsuya Mutou Tomotoshi Nishino, Manfred Sigrist, Matthias Troyer, 
and Hirokazu Tsunetsugu.  
We are also benefitted from discussions with Hiroshi
Kontani and Beat Ammon.  This work is financially supported by
Grant-in-Aid from the Ministry of Education, Science, Sports and 
Culture of Japan.


\begin{references}
\bibitem{Luttinger}
J.M. Luttinger, Phys. Rev. {\bf 119}, 1153 (1960).
\bibitem{Yosida}
K. Yosida, Phys. Rev. {\bf 147}, 223 (1966).
\bibitem{RKKY} 
M.A. Ruderman and C. Kittel, Phys. Rev. {\bf 96}, 99 (1954);
T. Kasuya, Prog. Theor. Phys. {\bf 16}, 45 (1956);
K. Yosida, Phys. Rev. {\bf 106}, 893 (1957).
\bibitem{RMP}
H. Tsunetsugu, M. Sigrist, and K. Ueda, Rev. Mod. Phys.
{\bf 69}, 809 (1997).
\bibitem{THUS}
H. Tsunetsugu, Y. Hatsugai, K. Ueda, and M. Sigrist, Phys. Rev. 
{\bf B46}, 3175 (1992).
\bibitem{DMRG}S.\ R.\ White,
Phys.\ Rev.\ Lett.\ $\bf 69$, 2863 (1992);
Phys.\ Rev.\ B $\bf 48$, 10 345 (1993).
\bibitem{shibata}
N. Shibata, J. Phys. Soc. Jpn, {\bf 66}, 2221 (1997).
\bibitem{wang}
X. Wang and T. Xiang, Phys. Rev. B{\bf 56}, 5061 (1997).
\bibitem{SATSU}
N. Shibata, B. Ammon, M. Troyer, M. Sigrist and K. Ueda,
J. Phys. Soc. Jpn {\bf 67}, 1086 (1998).
\bibitem{Ostlund}
S. Ostlund and S. Rommer, Phys. Rev. Lett. $\bf 75$, 3537 (1995).
\bibitem{Haldane}
F.D.M. Haldane, J.\ Phys.\ C $\bf 14$, 2585 (1981).
\bibitem{Schulz}
H.J. Schulz, Phys. Rev. Lett. {\bf 64}, 2831 (1990).
\bibitem{STU}
N. Shibata, A. Tsvelik, and K. Ueda, Phys. Rev. B{\bf 56},
330 (1997).
\bibitem{SUNI}
N.~Shibata, K.~Ueda, T.~Nishino, and C.~Ishii,
Phys.\ Rev.\ B $\bf 54$, 13495 (1996).
\bibitem{wire1}M.~Fabrizio and A.~O.~Gogolin,
Phys.\ Rev.\ B $\bf 51$, 17827 (1995).
\bibitem{wire2}R.~Egger and H.~Grabert,
Phys.\ Rev.\ Lett.\ $\bf 75$, 3505 (1995).
\bibitem{wire3}R.~Egger and H.~Schoeller,
Czech. J.\ Phys. $\bf 46$, Suppl.\ S4, 1909 (1996).
\bibitem{UNT}
K. Ueda, T. Nishino, and H. Tsunetsugu, Phys. Rev. B{\bf 50}, 
612 (1994)
\bibitem{fujimoto} S.\ Fujimoto and N.\ Kawakami,
J.\ Phys.\ Soc.\ Jpn.\ $\bf 63$, 4322 (1994).
\bibitem{whiteaff} S.~R.~White and I.~Affleck,
Phys.\ Rev.\ B $\bf 54$, 9862 (1996).
\bibitem{sikkema} A.~E.~Sikkema, I. Affleck, and S.~R.~White,
Phys. Rev. Lett. {\bf 79}, 929 (1997).
\bibitem{yamanaka} M. Yamanaka, M. Oshikawa, and I. Affleck,
Phys. Rev. Lett. {\bf 79}, 1110 (1997).
\bibitem{Kolomeisky}
E.B. Kolomeisky and J.P. Straley, Rev. Mod. Phys. {\bf 68},
175 (1996).
\bibitem{FoscHM}G. Bed\"{u}rftig, B. Brendel, H. Frahm, R.M. Noack, 
cond-mat/9805123 (1998).
\bibitem{Yu_White}C.\ C.\ Yu and S.\ R.\ White,
Phys.\ Rev.\ Lett.\ $\bf 71$, 3866 (1993).
\bibitem{Tsve}A.\ M.\ Tsvelik,
Phys.\ Rev.\ Lett.\ $\bf 72$, 1048 (1994).
\bibitem{Fujimoto2} S.\ Fujimoto and N.\ Kawakami,
J.\ Phys.\ Soc.\ Jpn.\ $\bf 66$, 2157 (1997).
\bibitem{MEM1}R. N. Silver, D. S. Sivia, and J. E. Gubernatis,
Phys. Rev. B $\bf 41$, 2380 (1990).
\bibitem{MEM2}J. E. Gubernatis, M. Jarrell, R. N. Silver,
Phys. Rev. B $\bf 44$, 6011 (1991).
\bibitem{MEM3}M. Jarrell and J. E. Gubernatis,
Phys. Rep. $\bf 269$, 133 (1996).
Phys.\ Rev.\ Lett.\ $\bf 72$, 1048 (1994).
\bibitem{Rice}T.\ M.\ Rice and K.\ Ueda,
Phys.\ Rev.\ Lett.\ $\bf 55$, 995 (1985);
Phys.\ Rev.\ B $\bf 34$, 6420 (1986).
\bibitem{Igarashi}J.\ Igarashi, T.\ Tonegawa, M.\ Kaburagi, and 
P.\ Fulde, Phys.\ Rev.\ B $\bf 51$, 5814 (1995).
\bibitem{Nishino}T.\ Nishino and K.\ Ueda,
Phys.\ Rev.\ B $\bf 47$, 12451 (1993).
\bibitem{Lieb}E.\ H.\ Lieb and F.\ Y.\ Wu,
Phys.\ Rev.\ Lett.\ $\bf 20$, 1445 (1968).
\bibitem{Mutou} T. Mutou, N. Shibata, and K. Ueda,
preprint (1998).

\end{references}
\end{document}